\documentclass[12pt,preprint]{aastex}  
\usepackage{epsfig}  
\usepackage{psfig}  
  
\usepackage{natbib}  
  
\begin{document}  
  
  
\title{Spectral Evidence for Widespread Galaxy Outflows at z$>$4}

  
\author{Brenda Frye\altaffilmark{1}}  
\affil{Department of Astrophysical Sciences, Princeton University, Peyton Hall, Princeton, NJ  08544}  
  
\author{Tom Broadhurst}  
\affil{Racah Institute of Physics, The Hebrew University, Jerusalem, Israel 91904}  
  
\and  
  
\author{Narciso Ben\'\i tez}  
\affil{Department of Physics and Astronomy, 
Johns Hopkins University, 3400 N. Charles Street, Baltimore, MD  21218-2686}  

\altaffiltext{1}{NSF Astronomy and Astrophysics Postdoctoral Fellow}

  
  
\begin{abstract}  
  
We present discovery spectra of a sample of eight lensed galaxies at
high redshift, $3.7<$z$<5.2$, selected by their red colors in the
fields of four massive clusters: A1689, A2219, A2390, and AC114. Metal
absorption lines are detected and observed to be blueshifted by
300-800 km s$^{-1}$ with respect to the centroid of Ly-$\alpha$
emission.  A correlation is found between this blueshift and the
equivalent width of the metal lines, which we interpret as a
broadening of saturated absorption lines caused by a dispersion in the
outflow velocity of interstellar gas. Local starburst galaxies show
similar behavior, associated with obvious gas outflows.  We also find
a trend of increasing equivalent width of Ly-$\alpha$ emission with
redshift, which may be a genuine evolutionary effect towards younger
stellar populations at high redshift with less developed stellar
continua.  No obvious emission is detected below the Lyman limit in
any of our spectra, nor in deep $U$ or $B$-band images. The UV
continua are reproduced well by early B-stars, although some dust
absorption would allow a fit to hotter stars.  If B-stars dominate,
then their relatively prominent stellar absorption lines should
separate in wavelength from those of the outflowing gas, requiring
more detailed spectroscopy.  After correcting for the lensing, we
derive small physical sizes for our objects, $\sim$ 0.5-5 kpc h$^{-1}$
for a flat cosmology with $\Omega_m=0.3, \Omega_{\Lambda}=0.7$.  The
lensed images are only marginally resolved in good seeing despite
their close proximity to the critical curve, where large arcs are
visible and hence high magnifications of up to $\sim20\times$ are
inferred. Two objects show a clear spatial extension of the
Ly-$\alpha$ emission relative to the continuum starlight, indicating a
``breakout'' of the gas.  The sizes of our galaxies together with
their large gas motion suggests that outflows of gas are common at
high redshift and associated with galaxy formation.
  
\end{abstract}  
  
  
  
  
\section{Introduction}  
  
  The photometric selection of high redshift galaxies by their
Lyman-limit break has proven to be a successful means of obtaining
galaxy redshifts in the range $2.5<$ z $<3.3$ \citep{Steidel:92,
Steidel:96a, Steidel:96b, Lowenthal:97}.  At somewhat higher redshift,
photometric selection is less reliable because the forest depression
of the continuum between Ly-$\alpha$ and the Lyman limit can be confused
with evolved or dusty galaxies at lower redshift. Moreover, these
Lyman-depressed galaxies at z$\sim$4 are significantly fainter than the
Ly-break galaxies, making spectroscopic work challenging. Despite
this, a sample of 48 $B$-band dropout selected galaxies has been
spectroscopically confirmed by \citet{Steidel:99}. These are observed with
relatively low spectral resolution, sufficient for determining
redshifts in the range $3.8<$ z $<4.5$, and occur with a frequency consistent with
little evolution relative to the lower redshift Ly-break galaxies
\citep{Steidel:99}.
  
Only a handful of higher redshift galaxies are known, the best 
examples being serendipitous discoveries boosted by lensing
\citep{Warren:96, Franx:97, Trager:97, Frye:98}.  Others are
identified primarily on the basis of an asymmetric emission line taken to be
Ly-$\alpha$ \citep{Dey:98}, or by virtue only of a relatively
large lower limit for the equivalent width of single emission lines
\citep{Weymann:98, Hu:98, Hu:99, Stern:00b, Rhoads:00, Manning:00}.  In
principle, high quality photometry alone is sufficient for z$>$4.5 candidates
because the strength of the Lyman-series forest depression \citep{Madau:95} 
can be established from bright QSO spectra \citep{Schneider:89,Storrie:96}, 
yielding precise predictions about the galaxy colors at those redshifts.
This continuum suppression feature has led to convincing claims of z$>$5 
for the very reddest objects in the Hubble Deep Field
\citep{Lanzetta:96,Spinrad:98}, and more recently, to the successful
selection of high redshift QSOs by \citet{Stern:00} and the Sloan Digital Sky 
Survey collaboration \citep[and references therein]{Fan:00,Anderson:01}.
    
Here we make use of this established high redshift Lyman-series depression
to color select galaxies for spectroscopy near the critical curves of
some of the most massive lensing clusters. This approach provides a
viable way of obtaining useful spectroscopy of high-z
galaxies, under favorable circumstances, where lensing is compounded by a
massive cluster galaxy lying close to the critical curve of a cluster
\citep{Franx:97,Frye:98}.
  
The galaxies presented here form part of a systematic redshift survey
of red-selected galaxies behind massive lensing clusters.  Most
galaxies selected in this way are early type lying at modest redshift
behind the cluster. We present the full redshift survey, and
measurements of cluster masses by magnification, in a different paper
\citep{Frye:01}.  Here we show only the most distant galaxies
discovered in our survey as a by-product of the red selection, and
whose optical colors place them in the range 3.5$<$z$<$6.5
\citep{Benitez:00}. We analyze the spectral information from 8 lensed
galaxies detected with redshifts of z~$=3.77, 4.04, 4.07, 4.25, 4.46,
4.67, 4.87,$ and $5.12$. These objects are relatively bright, allowing
for detailed spectroscopic work, and are suitable for follow-up at
other wavelengths.
  
This paper is organized as follows.  In \S2 we describe the target
selection and in \S3 we outline the photometric and spectroscopic
observations and data analysis techniques.  This is followed by a
presentation of the photometric and spectroscopic results in \S4, an
examination of the agreement with photometric redshift results in \S5,
and an analysis of gas motion and the equivalent width of the Ly-$\alpha$
emission line in \S6.  We close in \S7 with a discussion of our main
results.
  
\section{Target Selection}   
  
The lensing clusters were chosen to have the largest possible critical
radii and to lie at relatively low redshift so as to maximize the
number of background galaxies redder than the cluster early type
sequence.  Most of our color-selected galaxies have modest redshifts,
0.4$<$z$<$1.2 \citep{Frye:01}, but a small number turn out to be
forest-depressed high redshift galaxies.  Four clusters were used:
A1689 (z=0.18), A2390 (z=0.23), A2219 (z=0.23), and AC114
(z=0.31). These are massive, X-ray luminous lensing clusters which
show giant arcs and examples of multiply lensed images
(Figs.~\ref{fig:a1689id}, \ref{fig:4p04}, \ref{fig:a2219id}).
  
Target selection for spectroscopy is based on color and not on
elongated morphologies.  This is because in general relatively
circular mass substructure can create apparently undistorted images
({\it e.g.}~\citet{Trager:97}).  Also, in the case of intrinsically
small sources, the elongation by lensing may still leave an object
poorly resolved even with Hubble Space Telescope (HST) resolution (for
this reason lensed QSOs do not appear arc shaped).  This is
demonstrated, for example, in Fig~\ref{fig:4p88}, which shows the
galaxy at z=4.868 along with its ground based $UBVIJ$ images.  The
object lies close to a giant arc and is therefore highly magnified, by
$\sim$10$\times$, but is only marginally resolved spatially due to its
intrinsically small size.
  
The minimum color for our selection is set by the upper envelope of
early type cluster members, which is sharply defined and varies
between clusters as a result of the k-correction \citep{Kaiser:95}. At
z$>$4, Ly-$\alpha$ falls longward of the $V$ band so that a color
difference between $V$ and $I$ will be apparent. At z=4.5 the
continuum depression amounts to $\sim$2 magnitudes in $V$
(Fig.~\ref{fig:plotvi}).
  
\section{Observations}  
  
\subsection{Imaging and Photometry}  
  
The magnitudes quoted here are SExtractor MAGBEST magnitudes
\citep{Bertin:96}, or for the very compact objects, aperture magnitudes
corrected to `total' magnitudes using stellar aperture corrections.  We used
the following imaging data for making our photometric measurements and
target selections.

\subsubsection{A2219}
We have obtained deep observations on Keck I with the Low Resolution
Imaging Spectrometer (LRIS) in July, 1996 in good seeing
$\sim$0\farcs7 \citep{Oke:95}, totalling 20 min in $V$ and 10 min in
$I$ (Fig.~\ref{fig:a2219id}).  Two giant arcs frame the very large
cluster member just below and west of center, with many other arclets
seen about the bright central galaxies.

\subsubsection{A1689}  
We obtained deep $V$ and $I$ band observations at the European Space
Observatory (ESO) New Technology Telescope (NTT) in 1995.  In 1999 we
obtained new deep $V, R, I,$ and $Z$ data taken in average seeing of
0\farcs8 with LRIS at Keck II. The largest arc is shown in 
Fig~\ref{fig:a1689id}, subtending 30 degrees around the
cluster center.  Many other giant arcs are visible at this radius,
following a contour encompassing the bright central galaxies.  The
photometric redshift analysis presented in \S5 is based on these data
plus additional $U$ band observations with the Du Pont telescope at
Las Campanas, $B$ band observations with the Nordic Optical Telescope
at La Palma, and deep $J, H, K$ imaging taken at ESO NTT with the Son
of ISAAC (SOFI) IR imaging spectrograph. The reduction and analysis of
these $UBVRIZJHK$ data will be presented in an upcoming paper
\citep{Benitez:02}.

\subsubsection{A2390}
The imaging data were taken at the Canada France Hawaii Telescope in
$V$ and $I$, and for the core region, archival HST images were
analyzed in $V$ and $I$ \citep{Squires:96, Frye:98}.  The righthand
panel in Fig.~\ref{fig:4p04} and the color plate in \citet{Frye:98}
show the arcs at z=4.04 and other giant arcs in the center of this
cluster.

\subsubsection{AC114}
The imaging data were taken at the Anglo Australian Telescope,
consisting of approximately 3 hours of imaging in the $B$ and
$I$ bands. For this cluster, the color selection was based only on
these two bands.

\subsection{Spectroscopy}  
  
\subsubsection{A1689}

Some of the spectroscopic observations were carried out with LRIS on
the Keck I telescope in June 1997, totalling 2h 43min through a
1\farcs0 slit and in seeing of 0\farcs6-0\farcs8.  The 300 line/mm
grating was used blazed at 5000\AA, resulting in a resolution of 12\AA
\ at 6000\AA, determined from unblended sky lines. Adequate relative
spectrophotometry is obtained by taking the spectrum of an early type
cluster member at z$=0.187$, observed simultaneously through the
multislit mask, and comparing it to the standard empirical E/SO
spectrum of \citet{Kennicutt:92}. These observations resulted in the
detection of three high redshift galaxies at z=3.77, z=4.87, and z=5.12, shown in
Figs.~\ref{fig:3p9}, \ref{fig:4p88}, and \ref{fig:5.12} respectively.
  
More spectroscopic observations of these and other red selected
galaxies behind A1689 were made with LRIS on Keck II in March and
April 1999, totalling 2h 20min and 2h 05min respectively, through a
1\farcs0 slit, and in good seeing of $\sim$0\farcs8 on both occasions.
The 400 line/mm grating was used blazed at 6700\AA, resulting in a
resolution of 10\AA \ at 6000\AA, determined from unblended sky lines.
Relative spectrophotometry for the z=4.868 galaxy is obtained using
the same fluxing curve established for the June 1997 data.  Although
fairly rough, the fluxing curve is found to be smooth across the
throughput, leaving the spectral features unaffected and therefore
adequate for the purposes of our redshift survey.  This set of
observations resulted in additional observations for the objects at
z=3.770 and z=5.120.
  
In total, three high-z galaxies were discovered behind A1689, with
exposure times of 3h 40min (z=5.12), 4h 40min (z=4.87) and 2h
20min and 3h 40min (z=3.77).

\subsubsection{A2219}  
The spectroscopy for the three high-z galaxies behind A2219 was taken with
LRIS on Keck II in August 1998 in 5h 40min of exposure with the 400
line grating (Figs.~\ref{fig:4p07}, \ref{fig:4p45}, \ref{fig:4p67}).
The slit width was 1\arcsec, resulting in a resolution of 10\AA \ and
a dispersion of 1.8 \AA/pix.  We observed the spectrophotometric
standard G24-9 under photometric conditions, providing a relative flux
calibration.
  
\subsubsection{A2390}
The spectroscopy for the one dimensional spectrum for the z=4.039
galaxy behind A2390 (Fig.~\ref{fig:4p04}) was taken with LRIS on Keck
I in June 1997 in 2hr 50min of exposure, and is described in a
previous paper \citep{Frye:98}. Further longslit spectroscopy at
higher spectral resolution was also obtained with Keck I in June 1997,
whose two dimensional spectrum is shown in the lower panel of
Fig~\ref{fig:4p04}.
  
\subsubsection{AC114}
The spectroscopy for the high-z galaxy at z=4.248 behind AC114 was
taken with LRIS on Keck II in August 1997 in 2hr 8min of exposure with
the 300 line grating blazed at 5000\AA.  The slit width was 1\arcsec,
resulting in a resolution of 12\AA \ at the blazing angle, determined
from unblended sky lines.
  
\subsection{Reductions}  
  
We developed our own IDL reduction software to handle the large number
of spectra efficiently ($\sim$40 targets per multislit mask with 3-6
exposures per mask), and to address issues relating to pronounced
flexure and other spectral image artifacts (see \S3.3.2 and \S3.3.3
below). Our aim is to maximize the signal-to-noise without resampling
the data, so that groups of pixels carrying faint continuum signal
have every chance of being detected as a coherent pattern in the final
reduced image.  Although purpose built, the code is general and can be
applied to single and multislit data.  It is also fast and convenient,
processing the raw fits files directly.
  
\subsubsection{Reduction Procedure}  

The reduction procedure begins by flattening the data via dome or
superflats, as available.  White light dispersed flatfields or `object
flats' are used to take out the wavelength dependent effects of the
pixel response.  Cosmic ray events are removed by $\sigma$-clipping
above a suitable threshold which depends mainly on the number of
independent exposures.  Fringing changes between the exposures, and
therefore is not removed completely by the dome flat, requiring
dithering the spectra along the slit to average away this problem at
long wavelengths ($>7500$\AA).

Spatial curvature corrections are important for tracking
the object plus background accurately, amounting to up to 5 pixels
from center to end of a slitmask (see \S3.3.2).  Its proper
characterization allows for the detection of the faintest objects,
where the continuum signal is not readily identified before
one dimensional extraction but can be located blindly with knowledge
of the spatial curvature plus the known device coordinates of the mask
and targets.  We identify cosmic rays incident on the data frames by
flagging the hottest members in each spatial column with the median,
and then by fitting low order polynomials and applying iterative
$\sigma$-clipping.  The mean value of the good background pixels is
then subtracted from the image.
  
The spectra typically have a relatively limited coverage of sky, with
on average only 10-40 pixels in a given spatial column with which to
determine the best estimate.  In these cases, and on
account of the spectral tilt, sky subtraction is achieved by
correcting for a gradient in the spatial direction of the sky values
(see \S3.3.2 below).  This distortion results in a slowly changing
gradient in sky values in the spatial direction which we fit with a
low order polynomial.  We find that since many more independent sky
pixels are averaged over in this way than in the standard procedure
for sky subtraction this leads to a more accurate correction,
eliminates the need for rebinning, and produces higher signal-to-noise
results.

The wavelength calibration is obtained accurately from the many bright
sky lines in the red, and the choice of lines is interactively improved
on with a cursor connection.  The 1d spectra are then extracted
without resampling, and following the spatial curvature.  Cosmic rays on
the object are flagged in one dimension, by $\sigma$-clipping
relative to the local continuum, and then removed across a stack of
exposures by interactive $\sigma$-clipping.  Thresholds are set for
acceptable number of cosmic `hits' per pixel stack, to avoid removal
of real spectral features.
  
\subsubsection{Curvature Correction}  
  
\subsubsubsection{Following the Curvature}

The data are distorted in the spatial direction with respect to the 2d
detector frame. We find that we can remove this 2d curvature without
repixelization by following the curvature of the initially straight
slit boundaries which lie between the multislit spectra on the
slitmask.  At each boundary the pixel values abruptly drop from the
sky value of typically $\sim$200 counts per pixel down to zero,
providing a well defined curvature.  We fit it with a low order
polynomial and then shift the solution onto the object spectrum.  We
then sum up the object signal in the dispersion direction, shifting
the entire pixel stack along spatially by on average 3-5 pixels from
center to end of the mask, as determined by the polynomial.  In this
way we avoid rectifying the 2d image or introducing noise into the
background-limited data.  Note for its ability to follow
background-limited objects over the full dispersion range, we advise
this approach over the conventional one of following the curvature of
the object itself.
  
\subsubsubsection{Spectral Tilt}

While good flatfielding may average out the background sky, it does
not account for a slow wavelength variation in the spatial direction.
This type of curvature imposes a tilt on the spectrum which is most
pronounced on the edges of the bright sky lines.  Here the large change
in signal from the background value to emission peak introduces an
increment-decrement pattern, resulting in an artificial P-Cygni type
profile that depends slowly on wavelength.  In between the sky lines
this effect can be traced well with a slowly varying polynomial.
Operationally, we fit a linear slope in the spatial direction to each
spatial pixel stack which is determined by smoothing over a running
boxcar along the dispersion direction to get a good local estimate.
This procedure of treating the sky as a smoothly varying 2D surface is
found to improve the signal-to-noise of the data considerably for
sections lying between the sky lines.

For bright sky lines, however, the slope in the spatial direction is
determined for each column and then subtracted, since the gradient of
the sky values is very steep and depends sensitively on where the
sky line lies with respect to the fixed pixel grid. Although only one
column is used at a time, this estimate is a fair one because the
sky lines are relatively bright.  Note that this process does not
require rebinning of the data, unlike the conventional approach of
rectifying the entire 2D frame with arc-lamp calibrations.

\subsubsection{Artifacts}  
  
Both punched and milled masks produce physical defects along the slits,
imprinting dark and bright spatially-dependent striations into the
data. The resulting artifacts of this nonuniform illumination
typically affect the background by $3$-$5$\%, enough to lose
background-limited spectra completely for an unlucky alignment where
an object falls into a trough.  The effect is much worse for the
punched masks, which were used in about one third of the experimental
runs.  The effect can be corrected for, to first order, by dividing
the object flats into the data, but, as with the fringing correction,
small time dependent changes occur that limit the usefulness of the
flatfield.
  
The spatial profile of the slit, or slit function, is complex, varying
on subpixel scales. The sampling of the slit function in the spatial
direction varies along the spectrum due to the curvature, and therefore
offers in principle the opportunity to define this function over
different pixel phases.  In this way one potentially can piece together
a well sampled function and apply it to the data.  Although this
algorithm is not part of the standard reduction code, it has been used
in some of the more gross examples of non-uniform illumination.
  
\section{Results}  
  
A total of eight high redshift galaxies were discovered as a
consequence of our $V-I$ color selection and these are listed in
Table~\ref{tbl-1},~\ref{tbl-2}.  details.  Table~\ref{tbl-1} gives the
imaging details in five columns: cluster name; position in J2000
coordinates; $I_{AB}$ magnitude and its 1$\sigma$ error; $(V-I)_{AB}$
color; and lensing corrected size, in kpc h$^{-1}$ for a flat
cosmology with $\Omega_m=0.3,
\Omega_{\Lambda}=0.7$. Table~\ref{tbl-2} gives the spectral details in six columns:  cluster name,
redshift (not heliocentric corrected); rest equivalent width of
Ly-$\alpha$; summed rest equivalent width of OI$\lambda$1302 \ +
SiII$\lambda$1304; summed rest equivalent width of SiII$\lambda$1260 +
OI$\lambda$1302 \ + SiII$\lambda$1304 + CII$\lambda$1336; and velocity
shift between the centroid position of Ly-$\alpha$ and the absorption
lines.
  
The apparent magnitudes of these objects are relatively bright owing
to the large magnifications, 23$<$$I_{AB}$$<$25, producing relatively
high quality spectra (Figs.~\ref{fig:3p9}, \ref{fig:4p88}, \ref{fig:5p1},
\ref{fig:4p04}, \ref{fig:4p07}, \ref{fig:4p45},
\ref{fig:4p67}, and \ref{fig:4p25}).  The redshifts are clearly all high,
for which the Lyman series redshifts into the $V$ band, suppressing
the flux in $V$ relative to $I$ and giving these high-z galaxies their
characteristically large colors of $V_{AB}-I_{AB} \ge 1.5$
(Fig.~\ref{fig:plotvi}). In addition, the $J$-band image detects our
brightest high redshift example easily (Fig.~\ref{fig:4p88}).
  
The results are organized as follows.  In \S4.1 we describe the
continuum models generated for comparison with the data, in \S4.2 we
give the error estimation, in \S4.3 we give the prescription for
estimating the magnification, and in \S4.4 we present the spectra and
images.

\subsection{Continuum Modelling}  
  
Stellar continuum fits are compared with the fluxed spectra.  The
model is composed of a single stellar spectral type, which we find is
sufficient for estimating the continuum shapes and determining the
relative line velocities.  Continuum light in the UV sampled at high
redshift is of course dominated by the hottest stars for any
reasonable initial mass function (IMF), given the steep dependence of
luminosity on temperature for opaque material \citep{deMello:00}.  A
more detailed fit to our data, using spectral synthesis models,
will be presented elsewhere.  Kurucz models with solar metallicities
were used to estimate the temperature of the dominant contribution to
the UV at restframe $\sim$1000\AA, where we sample their spectra
\citep{Kurucz:93}.  Reddening is expected to be important at short
wavelengths, as is underscored by the presence of metal lines in all
of the high redshift galaxies in our sample. The effects of reddening
and/or age variations on the colors and continuum shapes may be
constrained with the help of infrared information.

We include sources of opacity into the models as necessary.  The
requisite Lyman-series opacity, as defined by \citet{Haardt:96}, is
found to match the data well (Fig.~\ref{fig:4p88}).  Our highest
signal-to-noise data also require an HI opacity at the source, with
column densities of HI between $logN=21.1$ and $logN=21.5$
(Figs.~\ref{fig:4p07}, \ref{fig:4p45}, \ref{fig:4p67}).
  
The hot B-stars which typically give good fits to the
continuum also show an intrinsic Ly-$\alpha$ absorption which can
dilute an otherwise strong Ly-$\alpha$ emission.  This becomes
increasingly more pronounced for late B-stars, where the absorption
increases strongly with decreasing temperature.  Fig.~\ref{fig:4p67}
shows one of our model fits to the data.  Here we fit a simple
two component model consisting of a B3 star plus Lyman-series opacity.
Continuum fitting in this detail has not been possible for comparably
high redshift galaxies \citep{Dey:98, Hu:98}, for which the continuum
is only marginally detected at best, or in the case of the z$=4.92$ lensed
galaxy of \citet{Franx:97}, its continuum is contaminated by light
from a nearby cluster member.

\subsection{Error Estimation} 

Error estimates on the equivalent width of Ly-$\alpha$ emission and its
wavelength centroid are made with respect to the above continuum
models, in an attempt to deal with a potentially large gradient in
the continuum light profile under this line, due to HI absorption local
to the galaxy and potentially from stellar atmospheres. Hence, there is 
most likely a sizable systematic uncertainty in the equivalent width of
this line and also some smaller error on the central velocity. The
uncertainty in the model then is the main source of error, which we have
included in Table \ref{tbl-2}.

As 
 the equivalent width is measured with respect to the bestfit model, the
 errors we quote on it are based on the uncertainty in
 the normalization of the bestfit model. A 1$\sigma$ estimate of the
 continuum height is determined simply by the absolute value of
 the dispersion of the continuum in each wavelength bin about the mean
 continuum level, excluding sky pixels and the ends of the
 spectrum. This dispersion measure is then corrected downward by a
 factor representing the square-root of the number of useful
 wavelength bins over which the continuum fit is made. The equivalent width is
 then remeasured, shifting the normalization up and down by this error
 in normalization. For the absorption lines, the continuum height is
 better determined locally, as the continuum gradient is rather flat
 in general for these lines and so the standard manual technique of
 continuum and line selected spectral regions is adopted and the error
 again is assumed to be dominated by the uncertainty on the continuum
 fit.

The centroid velocity is less sensitive to the choice of model
and is determined simply from an intensity weighted first moment
of the wavelength integrated over a restricted range of wavelength
which is somewhat subjectively determined by the quality of the data 
but covering typically not much more than about twice the spectral 
resolution of $\sim 10$\AA. In most cases the emission line 
is not evidently well resolved and hence the central wavelength 
largely reflects the instrumental resolution and over sampling
of our data by a factor of about 6 for each wavelength bin 
relative to the resolution for the 400-line grating.

For the absorption lines, the centroid velocity is determined using a
local fit to the continuum height and, in the usual way, the
flux-weighted wavelength centroid is determined relative to the local
continuum level. The error on the velocity centroid is close to the
sampling limit of 1.8\AA \ or $\approx 70$ km s$^{-1}$ for the best
defined lines.  A purely empirically determined estimate can also be
obtained by comparing the redshifts from the different absorption
lines.  In this case the dispersion is somewhat larger than the
sampling, $\approx 110$ km s$^{-1}$, on average, providing an average
internal estimate. The larger of these two numbers we quote for
$\Delta V$ in Table \ref{tbl-2}, where the dispersion is determined
for each spectrum individually in the case where more than one
absorption line redshift can be measured.

\subsection{Magnification Estimation}  
  
 The main uncertainty in estimating the magnification of an image is
the gradient of the lensing mass profile. The flatter the profile, for
a fixed Einstein ring radius $\theta_E$, the greater the
magnification.  For a power law mass density profile $\Sigma \propto
\theta^{-p}$, where $p$ is the power law index, we may expect the
slope to lie in the range $0.5<p<1$, bracketing the projected singular
isothermal profile and the Navarro, Frenk, and White (NFW) slope near
the Einstein ring thought to be appropriate for a large cold dark
matter halo \citep{Navarro:95}.
  
It is relatively straightforward to obtain a useful expression for the
magnification at angular position $\theta$, $\mu(\theta)$, in terms of
$\theta_E/\theta$:
  
$$ 
\mu(\theta)^{-1}=
\Bigl[1-{\Bigl({\theta_E\over \theta}\Bigr)}^p\Bigr]
\Bigl[1-(1-p){\Bigl({\theta_E\over \theta}\Bigr)}^p\Bigr] 
$$  
  
For highly magnified images close to the Einstein radius ($\theta_e\approx\theta$),
this reduces to

$$
\mu(\theta) \approx {1\over{p^2}}{1\over{(1-\theta_E/\theta)}}
$$ 

In terms of the magnification by a singular isothermal lens, it
becomes $\mu(\theta) \approx{1\over{p^2}}\mu^{iso}$. Thus, without a
precise determination of the central slope $p$ it is not possible to
obtain an accurate estimate of the magnification.
  
This magnification correction can be applied to the fluxes and to the
sizes of the spatially resolved objects.  Note in the cases where only
the lensed major axis is resolved, as is typical of the objects in our
sample, then one measures the magnification via the tangential stretch
factor, $\mu_t$.  This is just the first part of the above product for
$\mu(\theta)^{-1}$, and is hence a smaller correction yielding a lower
limit on the magnification:
  
$$ 
\mu_t(\theta)^{-1}=
\Bigl[1-{\Bigl({\theta_E\over \theta}\Bigr)}^p\Bigr]  
$$
  
Close to the Einstein ring this becomes,
$$
\mu_t \approx {1\over{p}} \Bigl({1\over{1-\theta_E/\theta}}\Bigr)
$$
In terms of the tangential stretch factor for the isothermal case, $
\mu_t \approx {1\over{p}} \mu_t^{iso}$.  Hence, this correction is
inversely dependent on the slope of the mass profile for images close
to the Einstein ring.
  
In the following section we provide fluxes corrected for lensing by
the factor $\mu(\theta)$, and intrinsic sizes corrected for lensing by
the factor $\mu_t(\theta)$, covering what we assume to be the range of
plausible central slopes of $0.5<p<1.0$.  Note that the Einstein
radius $\theta_E$ is safely assumed to lie very close to the angles
subtended by the main giant arcs seen in each cluster, on account of the low
redshifts of the clusters selected, z$\approx0.2$.  This is because
the distance ratio, $d_{ls}/d_s$, of source-lens separation is very
similar to the source distance and hence the bend angle of light has a
negligible dependence on redshift for our purposes.

\subsection{Individual High-z Detections}    
  
Each high-z detection will be discussed in turn below. In the main
panel of each figure we show the signal versus observed wavelength on
the lower axis and restframe wavelength on the upper axis.  Continuum
models are compared with the data, where relevant (see \S4.1).  The
sky spectrum is included at the bottom of the panel in order to
evaluate the reliability of the spectrum more clearly.  It varies
markedly with wavelength due to the presence of many bright sky lines
and fringing in the red.  We attach a postage stamp image of each
object in two or more of the $UBVIJ$ bands, and center the object in
the overlayed circle.  Magnifications are computed and used to
estimate the unlensed flux and intrinsic size (see \S4.3). Note that the
objects are typically only marginally resolved, resulting in upper
limits on the sizes and lower limits on the magnitudes.  We measure
sizes for a flat cosmology with $\Omega_m=0.3, \Omega_{\Lambda}=0.7$.
  
\subsubsection{A1689}

\subsubsubsection{z=3.770}  
  
The image and spectrum for this multiply lensed object are shown in
Figs.~\ref{fig:a1689id}, and \ref{fig:3p9} respectively.  A shallower,
blue spectrum appears in the lefthand panel and a deeper, higher
resolution one on the right.  The spectrum shows very weak or absent
Ly-$\alpha$ emission, for which we quote an upper limit on the
equivalent width of $\sim$3 \AA.  Both the Ly-$\alpha$ and Ly-$\beta$
breaks are detected, and several metal lines are clearly identified as
SiII$\lambda$1260, OI and SiII at $\lambda$1302 \ and $\lambda$1304 \
respectively, CII$\lambda$1336, and SiIV$\lambda1398$ and
$\lambda1402$, from which we determined the redshift.
  
This object is obviously elongated by lensing (see image in righthand
panels), forming a tangential arc of angular size 2.2\arcsec.  The
object lies about 15\arcsec \ outside the Einstein ring, resulting in
a tangential magnification of 6-12 and a flux magnification of 6-24
for the uncertain slope in the range $0.5<p<1$.  This yields an
intrinsic size of $\approx$1.5-2.5 kpc h$^{-1}$ and an unlensed
magnitude in the range $26<I_{AB}<27.5$ (see Table \ref{tbl-1}).
    
\subsubsubsection{z=4.868}  

The image of our brightest object, $I_{AB}=23.3$, is shown in
Fig.~\ref{fig:a1689id}, and its fluxed spectrum in
Fig.~\ref{fig:4p88}.  From the emission line, readily identified as
Ly-$\alpha$, a distinctive break on the blueward side and the
clear presence of an absorption line at 7396 \AA, identified as
SiII$\lambda1260$, we determined the redshift. Ly-$\alpha$
is redshifted by 360$\pm$ 70 km s$^{-1}$ from the metal line, and is
obviously asymmetric with a more extended wing on the red side.  Such
a profile shape is now established as a common property of all high-z
galaxies where Ly-$\alpha$ is seen and the data are of sufficient
quality \citep{Franx:97,Frye:98,Dey:98,Pettini:01}.
  
 A distinctive wedge-like step shortward of Ly-$\beta$ is observed, as
predicted at high redshift \citep{Madau:95}, and caused by the increasing
opacity of the Lyman-series forest for wavelengths greater than that
of Ly-$\beta$ at the emission redshift.  A good continuum fit is found
to a late type B0 star (see \S4.1).  This is considerably bluer than
the B3 and B5 best fits to our other high-z spectra 
(Figs.~\ref{fig:4p07}, \ref{fig:4p45}, \ref{fig:4p67}, and Fig.~2 in
\citet{Frye:98}).
  
No significant flux is seen in the spectrum below the Lyman limit and
no emission is detected in the $U$ and $B$ bands down to our 2$\sigma$
limiting magnitudes of $U_{AB}=27.3$ and $B_{AB}=26.7$ (see also
Fig.~\ref{fig:4p88}), indicating that no significant flux has escaped
below the Lyman limit, at least along our line of sight. A more
detailed determination of the escape fraction will be published
elsewhere.  A deficit of continuum flux below the Lyman limit is not
unexpected for any galaxy in which star formation is centrally
concentrated. Of course in addition, any modest column lower redshift
intervening forest cloud in the range $z>4.3$ could also lead to a
non-detection in the $U$ and $B$ bands.
  
This is the reddest object measured in the field, $(V-I)_{AB}>3.2$,
and almost redder than a passively evolving elliptical is expected to
reach at maximum, near z=1 (See Fig.~\ref{fig:plotvi}).  Our deep
ground based image resolves the object in $I$ in seeing of 0\farcs8,
and detects it at the 3$\sigma$ level in $J$.

The intrinsic size of the object can be determined roughly by its
position relative to the Einstein radius, which we take to be
$\theta_E=50$\arcsec \ given by the main arc
(Fig.~\ref{fig:a1689id}).  The lensed image lies $\sim$20\arcsec
\ outside this radius, resulting in a tangential stretch factor of 3-7
and a flux magnification of 3-14 for a slope in the range
$0.5<p<1$. This yields an intrinsic size of $\approx$1-2 kpc h$^{-1}$ 
and an unlensed magnitude in the range $24.7<I_{AB}<26.2$.

\subsubsubsection{z=5.120}  
  
The highest redshift object in our sample lies at z=5.120
(Fig.~\ref{fig:a1689id}) and its spectrum is shown in
Fig.~\ref{fig:5p1}.  From the strong asymmetric emission line detected
at 7430 \AA (see inset), and a distinctive break on the blueward side
we determined the redshift.  The line has an obviously asymmetric
profile and the largest observed equivalent width of our sample,
$W_{\alpha} \sim$ 30
\AA. Identified as Ly-$\alpha$, we tentatively find absorption at the
expected position of SiII$\lambda1260$ at the $\sim2.5\sigma$ level.
Our deep imaging provides a lower limit of $(V-I)_{AB} > 1.5$, with
$I_{AB}=25.2$ and a $2\sigma$ upper limit of $V_{AB}>$26.5.  Below the
galaxy and sky spectrum are the images in $U,B,V,I,$ and $J$.  The
object appears clearly only in the $I$ band, where it is faint and
unresolved.  No flux is seen in the spectrum below the Lyman limit.
  
The magnification of this object is hard to determine with any
precision because it lies so close to the critical curve, within
5\arcsec, where the magnification diverges.  It is interesting that
despite the very large magnification, which must be in excess of $\mu
> 10$, the object is at best only marginally spatially resolved,
meaning that its luminous part is intrinsically very small, of order
0\farcs7/$\mu$ or $<0.5$ kpc h$^{-1}$ for the flat case.  We estimate
an unlensed apparent magnitude of $I_{AB}>$ 27.5.

\subsection{A2390}  
\subsubsubsection{z=4.039}  
  
The image and fluxed spectrum for this multiply lensed object is shown in Fig.~\ref{fig:4p04} and is discussed in detail in
a previous paper \citep{Frye:98}.  The prominent features detected are
the Lyman-series and Ly-$\beta$ breaks, and an asymmetric
Ly-$\alpha$ emission line which is redshifted by 420 km s$^{-1}$ with
respect to the clearly detected interstellar absorption lines of SiII
1260$\lambda$, OI and SiII at $\lambda$1302 \ and $\lambda$1304 \
respectively, and CII$\lambda$1336.  Also SiIV$\lambda$1398 and
$\lambda1402$ are detected and shown in Fig. 2 of \citet{Frye:98}.
  
The righthand panel shows the HST WFPC 2 color image constructed in
the F606W and F814W passbands, from which we clearly distinguish four
spatially stretched image components and individual HII regions along
the 5\arcsec \ and 3\arcsec \ long N and S components respectively
(see Fig.~\ref{fig:4p04}).  In addition, there is cluster elliptical
in the center of the inset image which has been removed with a
deVaucouleur's profile to reveal the four image components of this
`extended' Einstein cross.

We have constructed a simple isothermal lens model which provides an
adequate fit to the data.  It consists of a dominant central cluster
potential modified by the cluster member galaxy that splits the images
\citep{Frye:98}.  The magnification is hard to constrain well since
the images lie so close to the critical curve where the magnification
diverges. The HST images allow a reasonable estimate of the size of
this object given by the width of the lensed images in the radial
direction which is little affected by lensing if the slope is close to
isothermal. The measured width is 0.2\arcsec \citep{Frye:98},
corresponding to an intrinsic size of $\sim$1 kpc h$^{-1}$ for the
flat case.

\subsection{A2219}  

\subsubsubsection{z=4.068}  
  
This object is labelled in Fig.~\ref{fig:a2219id} and its
fluxed spectrum shown in Fig.~\ref{fig:4p07}. The spectrum has
several strong and nearly saturated metal lines which are identified
as SiII$\lambda$1260, OI$\lambda$1302 SiII$\lambda$1304,
CII$\lambda$1336, and SiIV$\lambda1398$ and $\lambda1402$, from which
we determined the redshift.  There is a broad absorption which we take
to be absorption by the A-band at $\lambda7600$ \AA.  Ly-$\alpha$ is
seen in emission and is redshifted by 830 km s$^{-1}$ with respect to
the metal absorption lines, indicating a substantial amount of gas
outflow.  The best fitted model is for a B5 stellar population (solid
line) with Lyman-series forest attenuation along the line of sight and
an HI column at the source of $log N = 21.1$.  The Lyman-series break
appears shortward of Ly-$\alpha$ and no flux is detected below the
Lyman limit in the spectrum.  The $V$ and $I$ band images in the
righthand panels show the object to be faint, $I_{AB}=24.1$, and a
$V$-band dropout.  The magnification is likely to be small since the
object is at $\sim15$ critical radii, resulting in only a $\sim$10\%
correction to the observed magnitude and size.

 \subsubsubsection{z=4.445} 

The object at z=4.068 is labelled in Fig.~\ref{fig:a2219id} and its
fluxed spectrum is shown in Fig.~\ref{fig:4p45}.  It is a high quality
spectrum with a signal-to-noise of $\sim$7 per pixel in the continuum.
The most prominent features are the strong interstellar metal lines
and the sharp Ly-$\alpha$ break.  Ly-$\alpha$ is seen in emission and
has a characteristically asymmetric shape extending to the red.  This
is likely to be a dynamical effect as we discuss in \S6.1.  The
centroid of Ly-$\alpha$ is shifted redward by $\sim700$ km s$^{-1}$
with respect to the clearly identified lines of SiII$\lambda$1260,
OI$\lambda$1302, SiII$\lambda$1304, CII$\lambda$1336, and
SiIV$\lambda$1398 and $\lambda1402$, from which we determined
the redshift.  The best fitted model is for a B3 stellar population
(solid line) with Lyman-series forest attenuation along the line of
sight and an HI column at the source of $log N = 21.1$.  The object is
detected in both the $V$ and $I$ images shown to the right of the
spectrum.

The magnification of the image is modest, as it lies about 3 critical
radii from the center, resulting in a tangential stretch factor of
$<3$ and a flux magnification of $<$6 for the uncertain slope in the
range $0.5<p<1$.  The intrinsic size is estimated to be $\sim2-4$ kpc
h$^{-1}$ for the flat case, and its unlensed magnitude to be in the
range $24<I_{AB}< 25.5$.
 
\subsubsubsection{z=4.654} 

The object at z=4.654 is labelled in Fig.~\ref{fig:a2219id} and its
fluxed spectrum is shown in Fig.~\ref{fig:4p67}.  The signal to noise
is good, reaching $\approx$4 in the continuum at relatively high
spectral dispersion of 1.8 \AA/pixel.  Ly-$\alpha$ is in emission and
has a clearly asymmetric shape.  Shortward of this, a sharp break is
seen taken to be the Lyman-series break.  The interstellar metal line
SiII$\lambda$1260 is clearly detected in this spectrum, and other
lines commonly found in high-z galaxies are marked.  From the emission
line, continuum break, and absorption line we determined the
redshift.  Note that the broad absorption line at $\lambda7600$ \AA we
take to be absorption by the A-band.

Somewhat surprisingly, the stellar continuum rises redward of the
Ly-$\alpha$ emission line instead of declining as would be expected
from a hot stellar population with no absorption at the source.  This
feature is also seen in the slightly higher signal-to-noise spectrum
at z=4.445 in Fig.~\ref{fig:4p45}, and may be caused entirely by HI
absorption at the emission redshift or by the combined effect of HI
opacity and the stellar population.
  
The best fitted model is a B3 stellar population (solid line) which
has been attenuated by the Lyman-series forest and damped HI
absorption with column density $log N = 21.5$ placed at the emission
redshift.    There is a broad absorption which we take to be
absorption by the A-band at $\lambda7600$ \AA. The centroid of the
Ly-$\alpha$ emission line is shifted towards the red by 425 km
s$^{-1}$ with respect to this line, and has a total rest
equivalent-width of $W_{Ly\alpha} \sim 16$\AA.  The $V$ and $I$ band
images at the position of this arc are shown on the righthand side of
the spectrum with a flux of $I_{AB}=24.9$ and a color of
$(V-I)_{AB}>2.1$. The angular size in $I$ is small and unresolved.

The magnification is not large as this object lies at about 4 critical
radii, resulting in a tangential stretch factor of 1.5-3
and a flux magnification of 1.5-5 for the uncertain slope in the
range $0.5<p<1$.  The intrinsic size is estimated to be $<2$ kpc
h$^{-1}$ for the flat case, and the unlensed apparent magnitude,
$I_{AB}>25.2$.

\subsection{AC114}  
\subsubsubsection{4.248}  
  
The image and spectrum of the object at z=4.248 is shown in
Fig.~\ref{fig:4p25}. The spectrum shows a strong Lyman-series
absorption shortward of the position of Ly-$\alpha$, weak
Ly-$\alpha$ emission, and a marginal detection of the interstellar
line SiII$\lambda$1260 (2$\sigma$ level), from which we determined the
redshift.  The Ly-$\alpha$ emission line is redshifted by 460 km
s$^{-1}$ with respect to the metal line.  
The restframe equivalent width for
Ly-$\alpha$ has an upper limit of $W_{\alpha}=3$\AA.  The $B$ and $I$
band images in the righthand panels show that the object centered in
the circles is faint, $I_{AB}=23.9$, and a $B$-band dropout.  Note
that the bluer band for this object is $B_{AB}$ and not the usual
$V_{AB}$, as we did not obtain $V$-band data for this cluster.

This image lies about 20\arcsec \ outside of the Einstein radius,
resulting in a tangential stretch factor of $2.5-5$ and a flux
magnification of $2.5-10$ for the uncertain slope in the range
$0.5<p<1$.  The intrinsic size is estimated to be $<2$ kpc h$^{-1}$
for the flat case, and its unlensed magnitude to be in the range
$24.9<I_{AB}< 26.4$.

\section{Consistency with Photometric Redshifts}  
  
As described in \S3, we have obtained optical imaging in the $UBVRIZJH$ and $K$
bands for A1689, and very high quality photometric redshifts have been
determined for this field using the Bayesian method of
\citet{Benitez:00}. We compare the galaxy colors with those of a
forest opacity corrected redshifted template set, formed by the main
spectral types of \citet{Coleman:80}, plus two starbursts from
\citet{Kinney:96}.  We take into account the intrinsic color variation
or `cosmic variance' in the galaxy colors by convolving the resulting
color likelihoods with a gaussian of width $0.06(1+z)$, empirically
measured from the Hubble Deep Field (HDF) \citep{Benitez:00}. This
makes the range of final redshift probabilities considerably wider than they
would be if we included only the photometric error, but it yields a
more accurate representation of the true uncertainties involved in the
estimation.  Since we are working on a cluster field which strongly
magnifies the observed flux from background objects, we use the
redshift magnitude prior $p(z|T,m)$ introduced in \citet{Benitez:00}
for field galaxy distributions, but with the corresponding {\it
unlensed} magnitude, as calculated in \S4.

 The photometric redshifts estimated for the high-z objects in the
A1689 field are shown in Fig~\ref{fig:plotpz} and give excellent
agreement with the spectroscopic ones.  This is also true for the full
spectroscopic survey sample, and with this new photometry we have
identified even higher redshift targets out to z$\sim$6.5, for which
spectroscopic follow-up is being carried out with the Very Large
Telescope (VLT).
  
\section{Analysis}  
  
\subsection{Gas Motion}

\subsubsection{Velocity Dispersion}

In the spectrum of each object the metal line equivalent widths and
the velocity shift of the Ly-$\alpha$ centroid with respect to these
metal lines is measured (Table \ref{tbl-2}, Fig.~\ref{fig:8lya}).
There is an additional observation made for our object at z=4.04
\citep{Bunker:00}.  The analysis yielded an independent velocity shift
measurement which confirmed our original measurement of 420 km s$^{-1}$
\citep{Frye:98}, and so is not included separately here.
  
 Similar measurements from other high quality spectra are compiled
 from the literature for comparison
 \citep{Leitherer:96,Franx:97,Lowenthal:97,Kunth:98}.  The results are
 shown graphically for a local plus high-z sample
 (Fig.~\ref{fig:velocity}), and high-z sample only
 (Fig.~\ref{fig:velocity_all}).  The former includes only the two
 metal lines in common to all the samples, OI 1302$\lambda$ and SiII
 1304$\lambda$.  A fairly clear trend emerges in that the blueshift of
 the metal lines relative to the centroid of the Lyman-$\alpha$
 emission correlates with the equivalent width of metal line
 absorption.  Doppler broadening by the gas motion would leave the
 equivalent width unaltered for the optically thin case.  However, if
 the lines are optically thick, as we may expect for interstellar gas,
 then the width will be proportional to the velocity dispersion. The
 appearance of saturation is obvious in the spectra for the two
 objects in our survey at z=4.068 and z=4.445 in A2219
 (Fig.~\ref{fig:4p07} and
\ref{fig:4p45}).  In other cases it is not directly clear although our
limited spectral resolution is responsible for smoothing away the core
of lines of equivalent width less than the resolution of
$\sim$12\AA. Also we may expect that the spatial coverage of dense gas
does not fully cover the starlight continuum, so that the light which
is not absorbed partially fills the line.
  
 Hence, for saturated absorption lines, a spread in gas velocity
broadens the line and we propose that the trend we find in
Figs.~\ref{fig:velocity} and \ref{fig:velocity_all} is just a
consequence of an intrinsic spread in the dispersion in the gas
outflow velocity.  This in turn increases the equivalent width of the
absorption lines and the mean outflow velocity, thereby generating a
larger blueshift of the metal lines with respect to the systemic
velocity.  Note that the velocity width of the absorption lines is
only marginally resolved in the best cases with our limited spectral
resolution.

In addition to our data, blueshifts of the metal lines have also been
measured for the gravitational ring discovered by \citet{Warren:96},
in which the wavelengths of the IR emission lines have been compared
with Ly-$\alpha$. The same use of IR lines has been achieved for 15
relatively bright z$\sim$3 U-dropout galaxies by \citet{Pettini:01}.
They argue that restframe optical lines may better represent the
systemic velocity of the galaxy, rather than the resonantly scattered
Ly-$\alpha$ line.  Perhaps the cleanest measure of the mean expansion
rate is a comparison of IR lines with metal absorption lines, for
which most of their objects in that paper have measurements (their
Fig.~13).  The velocities lie in the range $+200$ to $-500$ km
s$^{-1}$, and are in fact very similar in magnitude to the velocity
offset between Ly-$\alpha$ and the metal lines.

\subsubsection{Models}

\citet{Legrand:97} convincingly explain the discrepancy in velocity of
the metal lines relative to Ly-$\alpha$ as arising from back scattered
Ly-$\alpha$ photons either emitted from the expanding HII region or
scattered from the inner `wall' of the surrounding HI, so that the
Doppler redshift of far side emission is sufficiently large to clear
it in wavelength of foreground absorption within the galaxy along the
line of sight. For the galaxies in which Ly-$\alpha$ emission is
absent, comprising about half the local starburst galaxies and distant
z$>2.5$ galaxies \citep{Steidel:96a,Steidel:96b,Steidel:99},
correspondingly strong Ly-$\alpha$ absorption is seen.  From this we
deduce that the most likely cause is a combination of a lower rate of
internal expansion and/or a higher HI column for absorption, rather
than simply absorption by dust.
  
In locally well studied examples of starburst galaxies with obvious
outflowing gas, \citet{Heckman:98} measure larger outflows for the
face on cases.  Since the galaxies at high redshift tend to be either
elongated by lensing or smaller than the seeing disk, it is difficult
to obtain inclination information.

\subsection{Equivalent Width of Ly-$\alpha$ Emission}  
  
We find a tentative trend in the equivalent width of Ly-$\alpha$
versus redshift across our range 3.7$<$z$<$5.2 (Fig.~\ref{fig:alpha}).
This is potentially a genuine evolutionary effect towards younger
stellar populations at high redshift at which time the stellar
continuum level will be lower because fewer stars have been created.
The time evolution of the equivalent width of Ly-$\alpha$ during a
star formation episode has been investigated in detail for single
burst models with a Salpeter IMF and a constant star formation rate
\citep{Bruzual:93,VallsG:98}.  Here a large equivalent width of
$W_{Ly\alpha}=250$ \AA \ is expected during the first $10^{6.5}$
years, when the stellar continuum builds up to a constant level given
by the lifetime of the hottest stars.  After this point, the
equivalent width declines, and after $10^7$ years is expected to be be
found in absorption due to the filling in by Ly$\alpha$ absorption
from B stars. Dust will of course reduce the ionizing flux further.
  
The main uncertainty on the estimate of the equivalent widths 
in faint sky dominated spectra arises from uncertainty 
in the continuum level underneath the emission line (see \S4). Our continuum
models are used to calculate this and show that depending on 
the stellar population and the degree of absorption close to
Ly-$\alpha$ this can be quite large. 

With all this in mind, if the formation rate of galaxies is
significantly increasing during our observed redshift interval, then
we might be witnessing genuine evolutionary behavior. On the other
hand, selection effects may be at work at some level.  We may have
missed galaxies which lie at the higher redshift end of our sample but
for which the equivalent width of Ly-$\alpha$ is low due the effects
of a small internal expansion rate or dust. Clearly this is a
potentially interesting result requiring larger samples of distant
galaxies for a fuller description.

\section{Discussion and Conclusions}  
   
  The claimed absence of Ly-$\alpha$ emission in deep narrowband
searches \citep{Thompson:95} was readily explained by dust absorption
along the hugely increased path length of resonantly scattered
Ly-$\alpha$ emission \citep{Charlot:93}, for even relatively modest
columns of neutral hydrogen. However, the recent discovery of copious
numbers of high redshift galaxies at 2.5$<$z$<$3.5, half of which
show Ly-$\alpha$ in emission \citep{Steidel:99,Rhoads:00}, revises this
conclusion. 

The explanation for the escape of Ly-$\alpha$ emission is
plausibly explained as a dynamical effect from outflowing gas, based
on detailed UV spectroscopy of nearby starburst galaxies
\citep{Legrand:97}. This outflow model, although not yet quantified,
can account for the peculiar observations reported here and for other
high quality spectra of high redshift galaxies.  The Ly-$\alpha$
photons generated by hot stars in the center of HII regions can escape
if back scattered from the inner ``wall'' of the surrounding expanding
gas, thereby avoiding the fate of the average Ly-$\alpha$ photon which
scatters and then is absorbed by dust within the surrounding
gas. These redder, back scattered photons take on the outflow
velocity.  If great enough, this velocity will exceed in wavelength
the redward wing of Ly-$\alpha$ absorption of the foreground gas through
which these photons must pass in order to reach the observer. In
contrast, the bluer photons will be preferentially absorbed, leading
to an asymmetric emission line.
  
 The internal gas motion implied by the mismatch of interstellar
absorption lines with emission lines of Ly-$\alpha$ and non-resonance
lines of OI detected in the IR has established, by analogy with local
starburst galaxies, that gas outflows are ubiquitous at early times.
If this gas escapes the potential of these early galaxies, then the
consequences for galaxy formation are predicted to be rather
important. 

The gas belonging to neighboring density perturbations
which have not yet virialized will be only tenuously held and be
expected to be stripped away by winds from nearby collapsed
starforming galaxies \citep{Scannapieco:01a}.  Since galaxy
formation is predicted to be much more spatially correlated at early
times by biasing \citep{Cole:89}, the effect locally is to
suppress the formation of small galaxies \citep{Scannapieco:01b}.  At
later times, this metal enriched gas is expected to become bound to
larger density perturbations which take longer to collapse and will be
heated and further enriched, modifying the cooling times of massive
galaxies and clusters \citep{Scannapieco:01b}.

The ubiquitous presence of metals in the IGM detected out to the
highest redshifts that the forest has been probed
\citep{Frye:93,Lu:96a,Lu:96b}, and the high level of enrichment of
cluster gas, may be most easily explained by a general pollution by
early galaxies \citep{Scannapieco:01a,Madau:01,Pettini:01} though
other mechanisms have been proposed to contribute, including tidal
stripping of processed gas \citep{Gnedin:97}, and photoevaporation of
gas in sufficiently small galaxies by UV background
\citep{Barkana:00b}.
  
Lensing helps to spatially resolve galaxies by stretching their images
so that one can obtain intrinsic size measurements.  We have made this
calculation for several of our highly magnified cases, which after the
correction for the likely level of magnification show that the objects
are in fact very small, $\sim$0.5-5~kpc h$^{-1}$. This is in reasonable
agreement with a sample of very distant red objects identified in the
HDF fields \citep{Spinrad:98, Weymann:98}.  It also agrees with
theoretical predictions \citep{Barkana:00a} for the early evolution
expected in the hierarchical models for structure formation, in which
the first galaxies to form are small, with increasingly more massive
objects collapsing and merging over time.
  
We conclude that the lensed galaxies presented here provide high  
quality information on the earliest known galaxies. Lensing has  
fortuitously afforded us detailed spectral and spatial details of  
magnified but presumably otherwise typical examples of galaxies at  
z$>$4.  The rarity of luminous high redshift galaxies and the  
requirement of a high magnification for useful spectroscopic follow-up  
means that such work will be slow but well rewarded with  
precious data on galaxies at otherwise inaccessibly early times.

\acknowledgments  
  
  We are grateful to Hy Spinrad for advice and to Richard Ellis, Garth
  Illingworth, Tim Heckman, and Alex Kim for helpful conversations.
  We thank Warrick Couch for valuable assistance in reducing the AC114
  data.  We thank Bev Oke and Judy Cohen for providing the LRIS
  instrument, and recognize the expert assistance of several of the
  staff at the Keck Observatories, including David Sprayberry, Randy
  Campbell, Bob Goodrich, Barbara Schaefer, Terri Stickel, Gary
  Puniwai, Ron Quick, and Wayne Wack, and Greg Wirth.  The W. M. Keck
  Observatory is a scientific partnership between the University of
  California and the California Institute of Technology, made possible
  by the generous gift of the W. M. Keck Foundation.  Finally, we
  thank the anonymous referee for a careful reading of the manuscript.

\bibliographystyle{/scr0/phd/astronat/apj/apj}  
  
{}

 
\clearpage  
  
  

\figcaption[plot_vi] {Color vs.~Redshift for the El, Scd and Im galaxy
types of \citet{Coleman:80}, the SB2 and SB3 starbursts from
\citet{Kinney:96}, and a B3 stellar model \citep{Kurucz:93}.  The six
galaxies in our sample with measured $(V-I)_{AB}$ color are plotted as
square shaped points with error bars, or as lower limits for the galaxies
at z=4.65, 4.87, and 5.12.  Our high-z sample follows the expected color
trend well.  Note the potential source for confusion with
elliptical galaxies at z$\approx1$. \label{fig:plotvi}}

\figcaption[plot_pz] {Bayesian redshift
probability distributions as defined in \citet{Benitez:00}, based
information from 9 optical and infrared bands in A1689.  The
photometric redshifts agree with the spectroscopically determined ones
to the 1$\sigma$ level.  There is a double peaked probability for the
galaxy at z=5.12, arising from its faintness and therefore relatively
large photometric errors.  Note, however, that the sum of the redshift
probabilities lies closest to the spectroscopic redshift.
\label{fig:plotpz}}
  
\figcaption[identify_a1689.eps]{Deep $I$-band image of the center of
A1689 (z=0.18) taken with LRIS, with the three high-z galaxies
discovered behind this cluster circled.  The lensing effect is very
clear, forming a general circular pattern of faint galaxies around the
center of the cluster.  Note the proximity of these objects to the
cluster critical curve and their small lensed sizes.  For reference,
north is up and east is to the left.  \label{fig:a1689id}}
  
\figcaption[z3p9.ps]{Blue and red spectra and images of the z=3.77 arc
behind A1689 (z=0.18), with the commonly found spectral features labelled.
Very weak Ly-$\alpha$ emission is detected with an upper limit on its
equivalent-width of $\sim$ 3\AA.  Both the Lyman-series and
Lyman-$\beta$ breaks are detected, as are the interstellar lines of
SiII$\lambda$1260, OI and SiII at $\lambda$1302 \ and $\lambda$1304 \
respectively, CII$\lambda$1336, and SiIV$\lambda1398$ and
$\lambda1402$.  The sky spectrum is shown in the lower attached
panels. Note this object's arc-like shape in $I$ and marginal
detection in $V$.  \label{fig:3p9}}
  
\figcaption[a1689_4p88_montage2]{Fluxed spectrum and images of the
z=4.868 galaxy behind A1689 (z=0.18), with common spectral features
marked.  Strong Lyman-series and Ly-$\beta$ breaks are seen.
Prominent Ly-$\alpha$ emission is detected which is asymmetric and
redshifted by 360 km s$^{-1}$ with respect to the clearly detected
SiII 1260$\lambda$ interstellar line.  The dashed curve shows the
bestfit unreddened B0 stellar population, and the thick solid curve,
the effect of attenuation by Lyman-series blanketing.  The sky
spectrum is shown in the lower attached panel.  Deep ground-based
images in $UBVI$ and $J$ used for color selection show it to be small
and only marginally resolved despite its proximity to the critical
curve and large inferred magnification.
\label{fig:4p88}}
  
\figcaption[a1689_5p1_montage2.ps]{Spectrum and images of the z$=5.12$
galaxy behind A1689 (z$=0.18$), with commonly found spectral features
marked.  A Lyman-series break is clearly detected together with a
prominent asymmetric emission line which we take to be Ly-$\alpha$
(see smoothed inset).  There is a marginally significant detection of
absorption at the expected position of SiII 1260$\lambda$, but at this
signal to noise is consistent with the continuum noise level. The sky
spectrum is shown in the lower attached panel.  Deep ground based
images in $UBVI$ and $J$ used for color selection show this $V$-band
dropout to be small and unresolved from the ground, despite its large
inferred magnification.  \label{fig:5p1}}
  
\figcaption[z404abc.ps]{One and two dimensional spectra and color
image of the quadruply lensed galaxy at z=$4.039$ behind A2390
(z=$0.23$).  The Lyman-series and Ly-$\beta$ breaks are clearly seen,
as well as a prominent and asymmetric Ly-$\alpha$ emission line which
is redshifted by 420 km s$^{-1}$ with respect to the clearly detected
interstellar absorption lines of SiII 1260$\lambda$, OI and SiII at
$\lambda$1302 \ and $\lambda$1304 \ respectively, and
CII$\lambda$1336.  The righthand panel shows the HST WFPC 2 color
image, with the N and S components extending spatially 5\arcsec \ and
3\arcsec \ respectively.  Note that the central elliptical galaxy has
been removed to reveal the four individual components of this
`extended' Einstein cross.  The elliptical is retained in the 2d
spectrum in the lower panel, shown as the bright swath in
between the spatially resolved N and S image components (labelled).
Note the faint continuum extending toward longer wavelengths (to the
right), and downward spatially from the bright Ly-$\alpha$ lines in
both the N and S components, demonstrating like image parity.
\label{fig:4p04}}
  
\figcaption[identify_a2219.eps]{Deep $I$ image of the center of A2219
(z=0.23) taken with LRIS, with the three high-z galaxies discovered at
z=4.068, z=4.445, and z=4.654 circled and labelled.  The massive
cluster members lie just below center and to the right.  This image
forms part of a multicolor set used for photometric target selection.
For reference, north is up and east is to the left.
\label{fig:a2219id}}
  
\figcaption[z4p07.ps]{Fluxed spectrum and images of the z=4.068 galaxy
behind A2219 (z=0.23).  A strong Lyman-series break is detected
against a bright stellar continuum.  Prominent Ly-$\alpha$ emission is
seen which is asymmetric and redshifted by 830 km s$^{-1}$ with
respect to the position of the several wide and nearly saturated
interstellar absorption lines of SiII 1260$\lambda$, OI and SiII at
$\lambda$1302 \ and $\lambda$1304 \ respectively, CII$\lambda$1336,
and SiIV$\lambda1398$ and $\lambda1402$.  The dashed curve shows the
bestfit unreddened B5 stellar population, and the thick solid curve,
the effect of attenuation by Lyman-series blanketing and absorption at
the position of Ly-$\alpha$ ($log N = 21.1$).  The sky spectrum is
included in the bottom attached panel.  Two deep $VI$ Keck images show
that this $V$-band dropout is only marginally resolved from the
ground. \label{fig:4p07}}
  
\figcaption[z4p45.ps]{Fluxed spectrum and images of the z$=4.445$
galaxy behind A2219 (z=0.23).  A strong Lyman-series break is detected
against a bright stellar continuum.  Prominent Ly-$\alpha$ emission is
seen which is asymmetric and redshifted by 700 km s$^{-1}$ with
respect to the position of the several clearly detected interstellar
absorption lines of SiII 1260$\lambda$, OI and SiII at $\lambda$1302 \
and $\lambda$1304 \ respectively, CII$\lambda$1336, and
SiIV$\lambda1398$ and $\lambda1402$.  The dashed curve shows the
bestfit unreddened B3 stellar population, and the thick solid curve,
the effect of attenuation by Lyman-series blanketing and absorption at
the position of Ly-$\alpha$ ($log N = 21.1$).  The 2d spectrum is
shown in the lower panel.  Note the spatially resolved Ly-$\alpha$
line, and continuum extending toward longer wavelengths (to the
right).  The sky spectrum is included in the bottom attached panel.
Two deep $VI$ Keck images show this $V$-band dropout to be resolved
but small. \label{fig:4p45}}
  
\figcaption[z4p67.ps]{Fluxed spectrum and images of the z$=4.654$
galaxy behind A2219 (z=0.23).  Strong Lyman-series and Ly-$\beta$
breaks are seen together with prominent Ly-$\alpha$ emission
redshifted by 425 km s$^{-1}$ with respect to the clearly detected
SiII 1260$\lambda$ interstellar line.  The dashed curve shows the
bestfit unreddened B3 stellar population, and the thick solid curve,
the effect of attenuation by Lyman-series blanketing and absorption at
the position of Ly-$\alpha$ ($log N = 21.5$).  The sky spectrum is
included in the bottom attached panel.  Two deep $VI$ Keck images show
that this $V$-band dropout is only marginally resolved from the
ground.  \label{fig:4p67}}  
  
\figcaption[z4p25.ps]{Spectrum and images of the z$=4.248$ galaxy
behind AC114 (z=0.31).  A strong Lyman-series break is detected
against a bright stellar continuum.  Ly-$\alpha$ is seen in emission
redshifted by 460 km s$^{-1}$ with respect to the position of the
interstellar absorption line SiII 1260$\lambda$, at the 2$\sigma$
level.  The sky spectrum is included in the bottom attached panel.
Two deep $BI$ images show this $B$-band dropout to be small and only
marginally resolved.  \label{fig:4p25}}

\figcaption[Lya_Spec_8b.eps]{The Ly-$\alpha$ emission line is plotted for
each of the high-z galaxies discovered (solid), along with the expected
location of the center of Ly-$\alpha$ emission based on the
redshift measured from the absorption lines (dashed). A systematic
bias is apparent which is interpreted as evidence for gas
outflow. Note that for the galazy at z=5.120, the continuum emission
is too weak for reliable absorption line detecton.  \label{fig:8lya}}

\figcaption[Velshift_OISiII.eps]{The blueshift of the metal lines
relative to the centroid of the Ly-$\alpha$ emission is plotted
against the summed rest equivalent width of two lines: OI 1302$\lambda$ and
SiIV 1304$\lambda$.  The data for the five emission line objects
discovered in our survey with measurable absorption line
equivalent widths are shown as circles and their redshifts marked.
The other symbols correspond to different sets of data for high-z and
local galaxies (see legend).  The increasing trend may be attributed
to a dispersion in gas outflow speed for saturated interstellar lines,
as discussed in the text.  \label{fig:velocity}}
  
\figcaption[Velshift_all.eps]{The blueshift of the metal lines
relative to the centroid of the Ly-$\alpha$ emission is plotted
against the summed rest equivalent width of several lines: SiII
1260$\lambda$, OI and SiII at $\lambda$1302 \ and $\lambda$1304 \
respectively, and CII$\lambda$1336.  The data for the five emission
line objects with measurable absorption line equivalent widths for all
of these lines are shown as circles and their redshifts labelled.  The
other symbols correspond to different high-z and local starburst
galaxies (see legend).  As in Fig.~\ref{fig:velocity}, the clear
increasing trend may be attributed to a dispersion in gas outflow
speed for saturated interstellar lines.  \label{fig:velocity_all}}
  
\figcaption[Lya_z.eps]{Rest equivalent width of Ly-$\alpha$ emission
as a function of redshift.  Although the scatter is large, a rough
increasing trend is seen, which may be a genuine evolutionary effect
towards younger stellar populations at high redshift with less
luminous stellar continua, although larger samples are required for a
more well constrained measurement.  \label{fig:alpha}}  
\clearpage

  
  

\begin{deluxetable}{ccccc}  
\tablewidth{5.5in}  
\tablecaption{Distant Galaxy Imaging Information \label{tbl-1}}  

\tablehead{

\colhead{Cluster} & \colhead{Position (J2000)}&$I_{AB} \pm 1\sigma$ & $(V-I)_{AB}$ & \colhead{Size}\\

 & & & & (kpc h$^{-1}$)  }
  
\startdata  
  
A1689\_1 & (13:11:29.980,-1:19:15.293) & $24.2 \pm 0.2$& 1.5 &2 \\ 

A2390\_1 & (21:53:33.669,17:42:04.184) &$23.0 \pm 0.3$ & & 1 \\ 

A2219\_1 & (16:40:12.638,46:44:59.412) &$24.1 \pm 0.2$ & 1.8 & 3.5 \\ 

AC114\_1 & (22:58:45.841,-34:49:33.806)
&$23.9 \pm 0.3$ &  & $<$2 \\ 

A2219\_2 & (16:40:15.885,46:43:58.696) & $23.7 \pm 0.2$& 1.8 & 3 \\ 

A2219\_3 & (16:40:11.152,46:41:58.343) &$24.9 \pm 0.2$ &
$>2.1$\tablenotemark{b}& $<$2 \\ 

A1689\_2 & (13:11:25.499,-1:20:51.900) & $23.3 \pm 0.1$ &
$>3.2$\tablenotemark{a} & 1.5 \\ 

A1689\_3 & (13:11:35.042,-1:19:51.617) &$25.0
\pm 0.2$ & $>1.5$\tablenotemark{a}& $<$0.5 \\ \enddata
  
\tablecomments{The columns are: Cluster; 
Object Coordinate (J2000); $I_{AB}$
magnitude with 1$\sigma$ error bars; $(V-I)_{AB}$ color; Intrinsic
Object Size, in kpc h$^{-1}$ for $\Omega=0.3$, and $\Lambda=0.7$.}
  
\tablenotetext{a}{2$\sigma$ limiting magnitude is $V_{AB}=26.5$}  
\tablenotetext{b}{2$\sigma$ limiting magnitude is $V_{AB}=27.0$}  
  
\end{deluxetable}  
\clearpage

\begin{deluxetable}{cccccc}  
\tablewidth{4.1in}
\tablecaption{Distant Galaxy Spectral Information \label{tbl-2}}  

\tablehead{    
  
\colhead{Cluster} & \colhead{z}& \colhead{$W_{\alpha}$} & \colhead{$W_{1}$} &  
\colhead{$W_{2}$} & \colhead{$\Delta V$} \\
  
 &$\pm0.001$ & \colhead{(\AA)} &\colhead{(\AA)}  & \colhead{(\AA)} &  
\colhead{(km s$^{-1}$)} }
  
\startdata

A1689\_1 & 3.770 & $\le3.3$ & 3.4 & 7.1 & \\

A2390\_1 & 4.039 & $8.84^{+5.5}_{-2.5}$ &1.3 & 6.1 & $420 \pm 100$ \\

A2219\_1 & 4.068 & $5.05^{+2.1}_{-2.2}$ &5.9 &13.8 & $830 \pm 100$ \\

AC114\_1 & 4.248 & $3^{+2.4}_{-1.6}$ & & &  \\

A2219\_2 & 4.445 & $9.05^{+4.6}_{-3.3}$ &3.8 &10.4 & $700 \pm 90$ \\

A2219\_3 & 4.654 & $10.4^{+6.8}_{-5.5}$ &1.8 & 5.7 & $425 \pm 120$ \\

A1689\_2 & 4.868 & $12.4^{+8.83}_{-3.84}$ &2.1 & 5.7 & $360 \pm 120$ \\

A1689\_3 & 5.120 & $29.7^{+13.29}_{-4.6}$ & & &  \\ 

\enddata
  
\tablecomments{The columns are: Cluster; Redshift (not
heliocentric corrected); Rest Equivalent Width of Ly-$\alpha$; Summed rest
equivalent width of OI 1302$\lambda$ + SiII 1304$\lambda$; Summed rest
equivalent width of SiII 1260$\lambda$ + OI 1302$\lambda$+ SiII
1304$\lambda$ + CII 1336$\lambda$; Velocity Shift between the
centroid position of Ly-$\alpha$ and the position of the interstellar
lines, with 1$\sigma$ error bars}
  
\end{deluxetable}  
\clearpage  
  
  

\begin{figure}[tp]  
\plotone{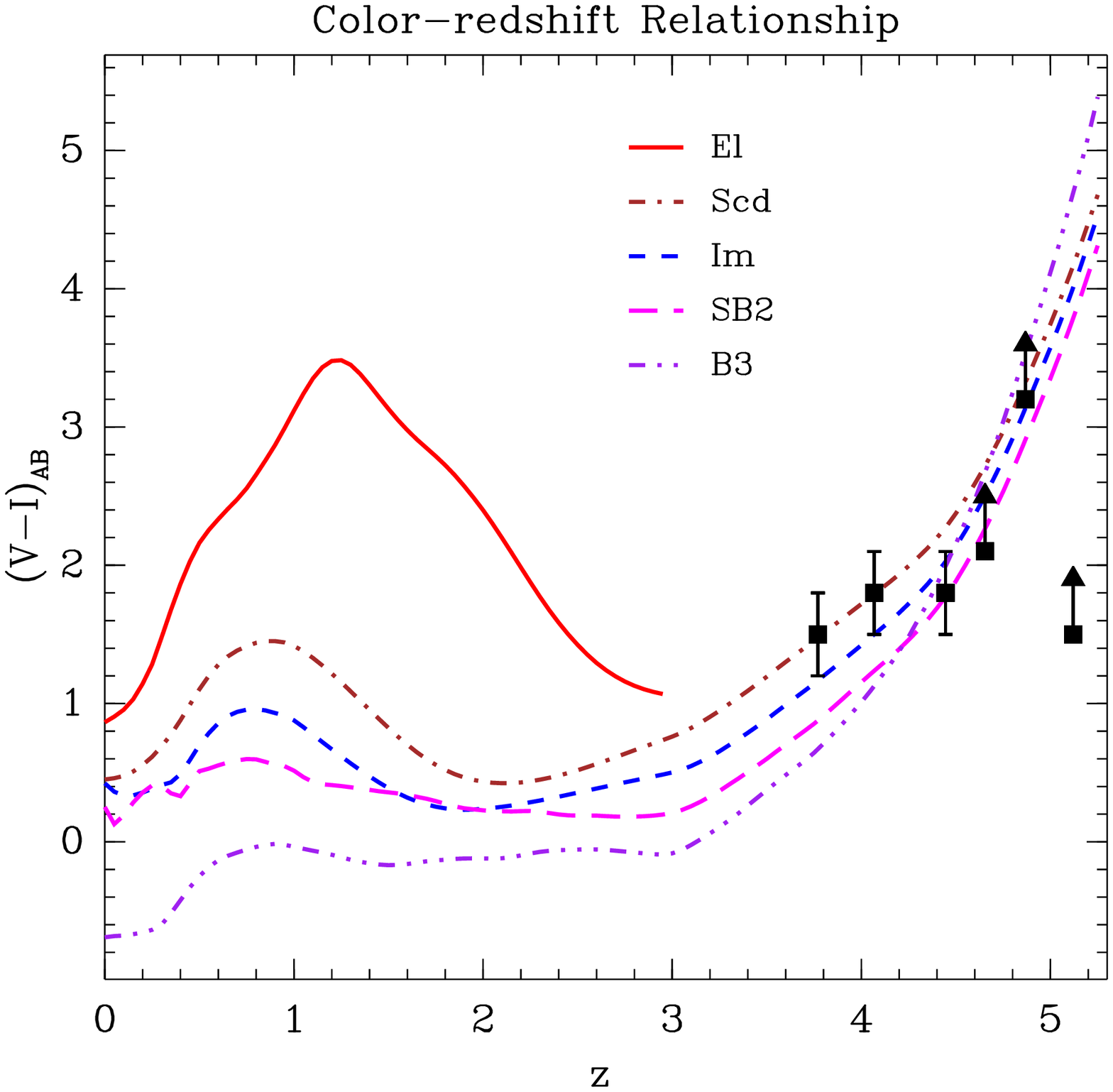}  
\end{figure}  
  
\clearpage  
  
\begin{figure}[tp]  
\plotone{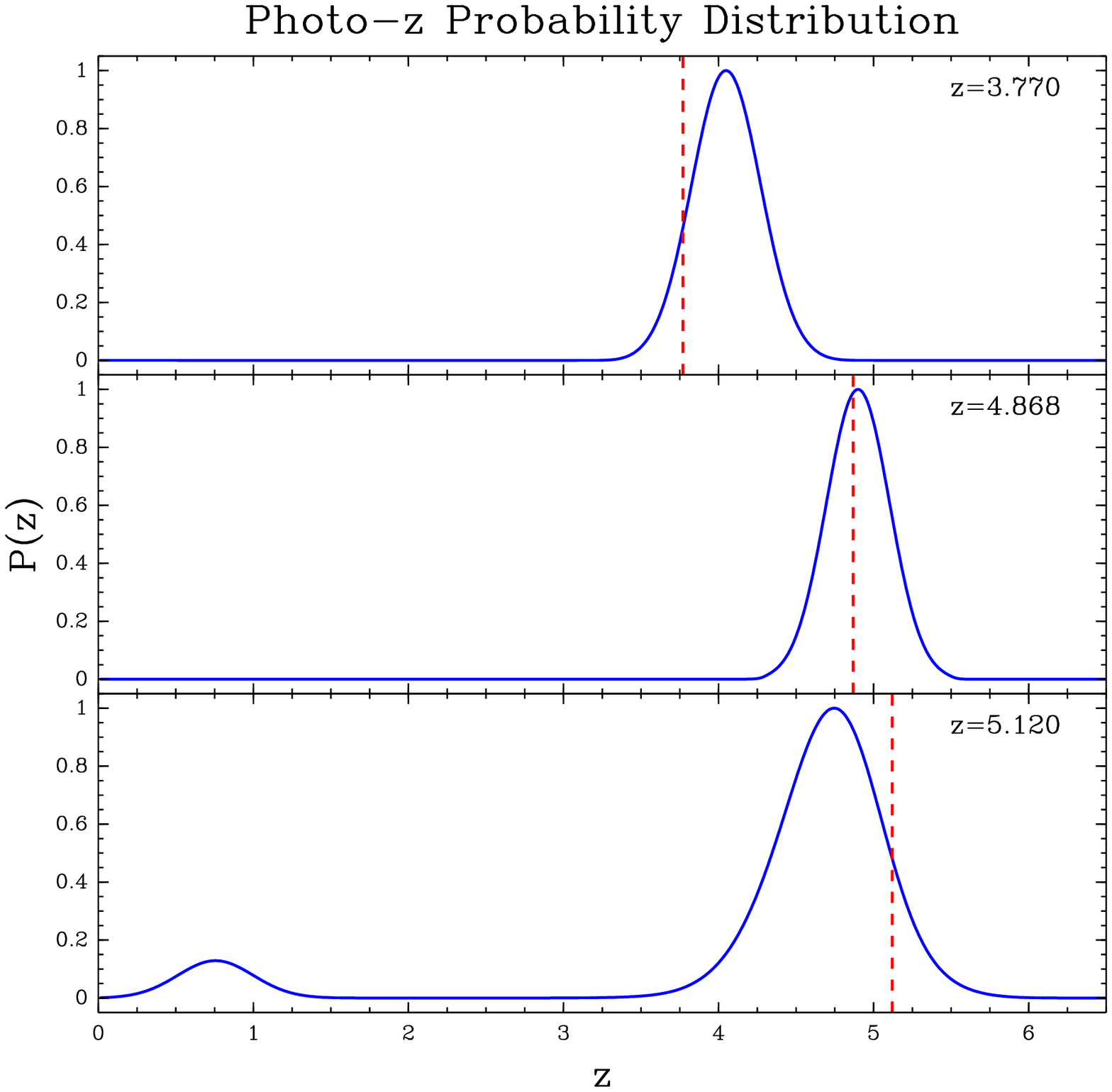}  
\end{figure}  
  
\clearpage  
  
\begin{figure}[tp]  
\plotone{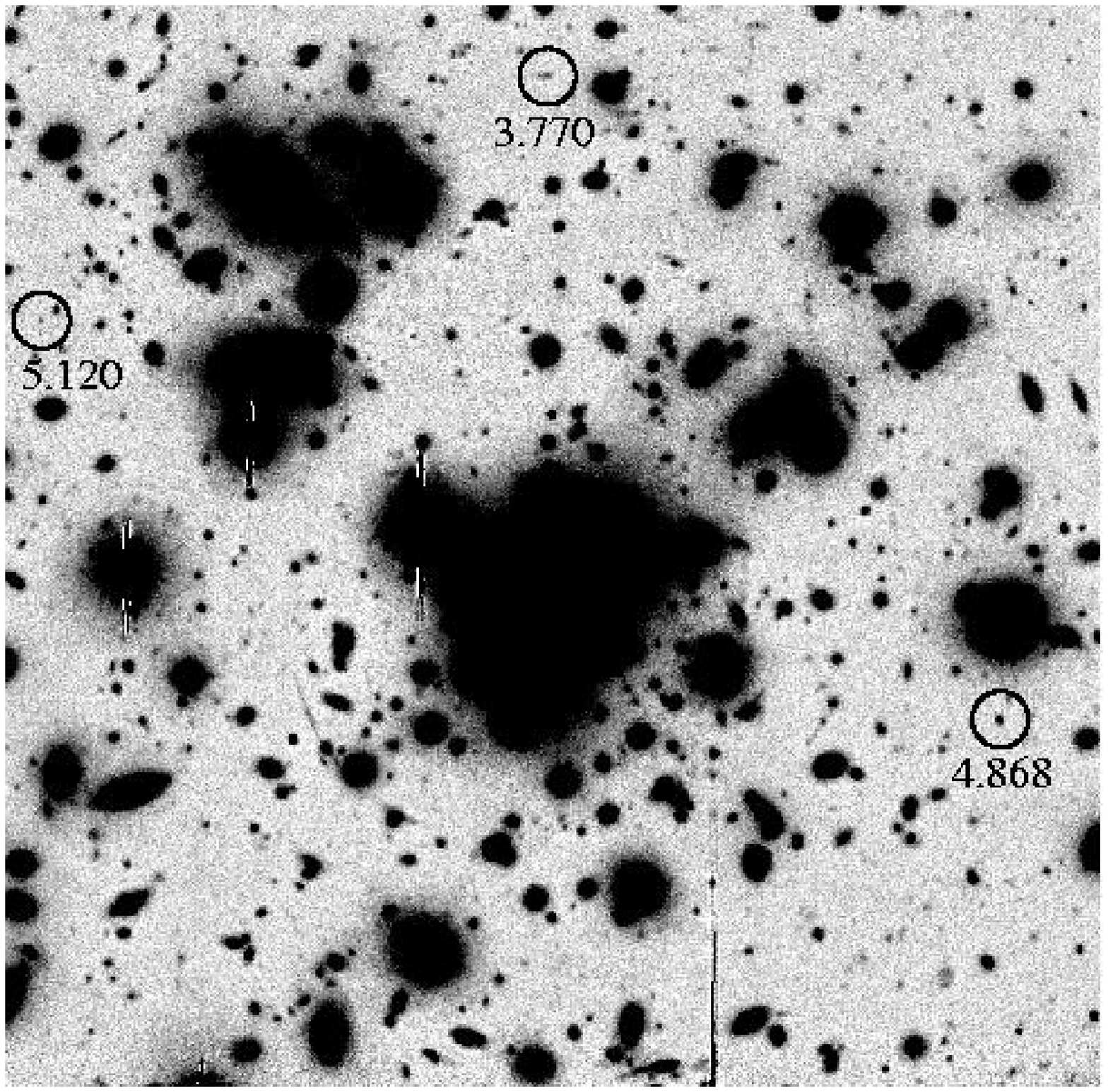}  
\end{figure}  
  
\clearpage  
  
\begin{figure}[pt]  
\plotone{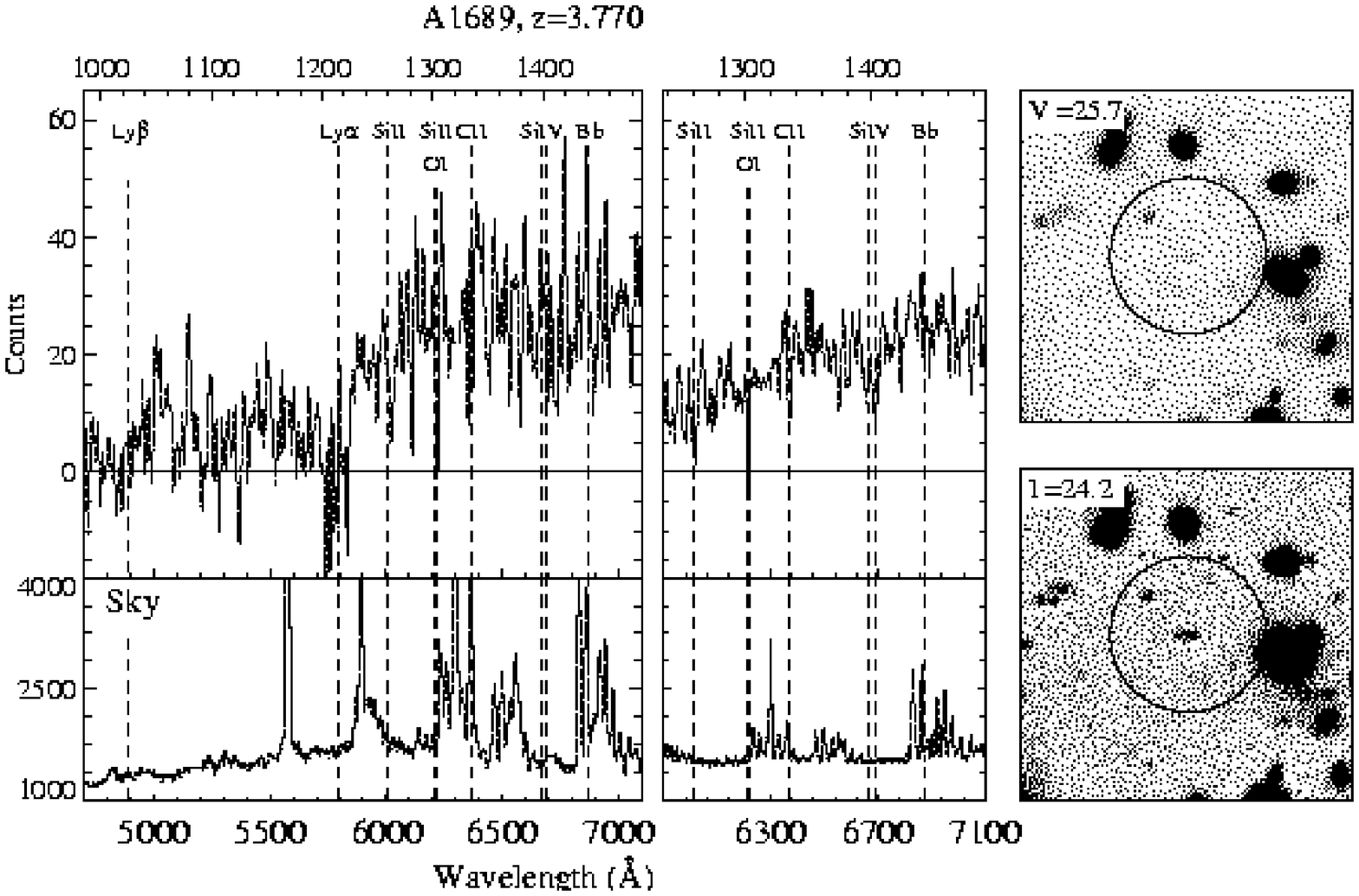}  
\end{figure}  
  
\clearpage  
  
\begin{figure}[tp]  
\epsscale{0.8}
\plotone{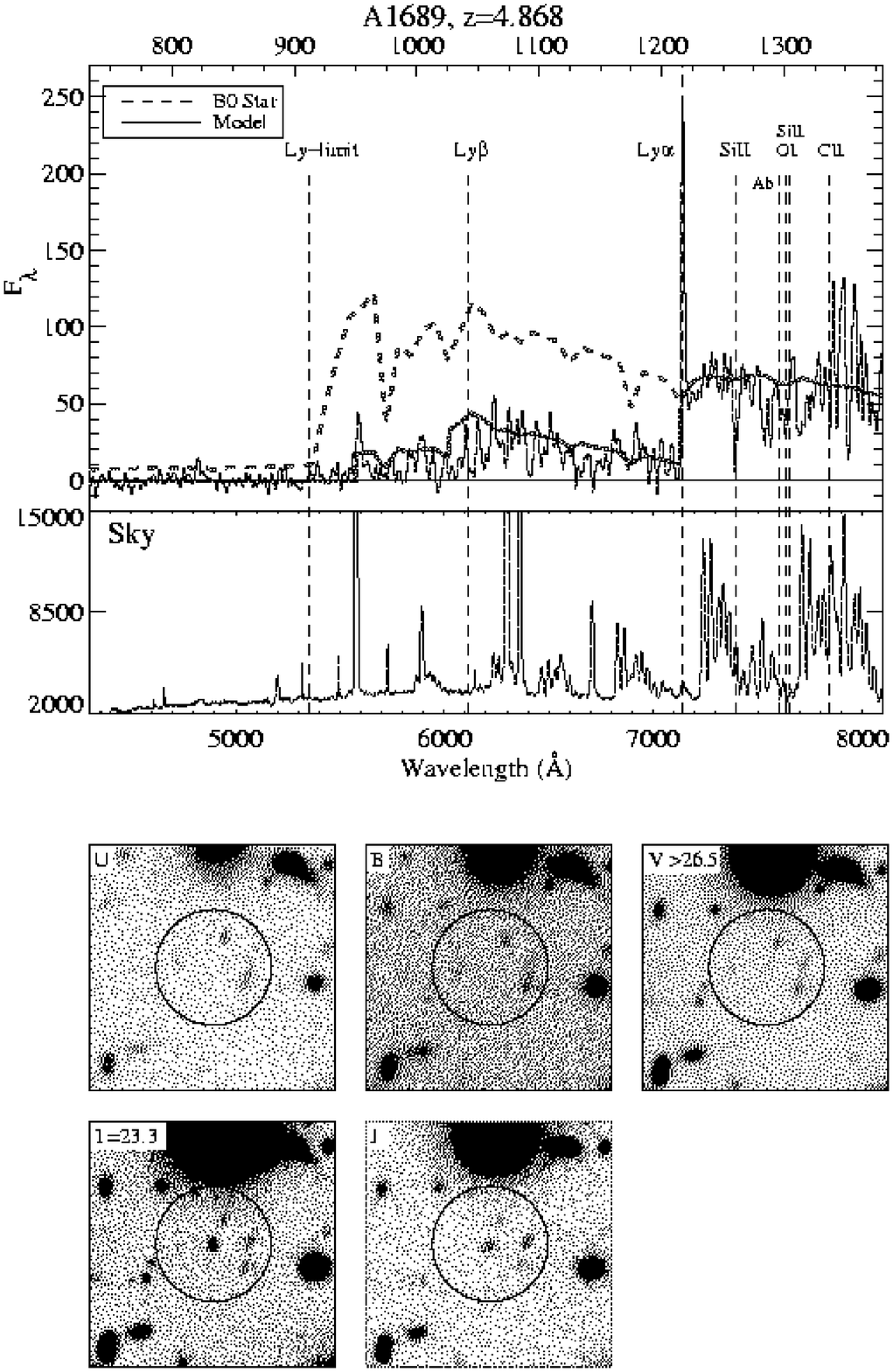}  
\end{figure}  
  
\clearpage  
  
\begin{figure}[pt]  
\epsscale{0.8}
\plotone{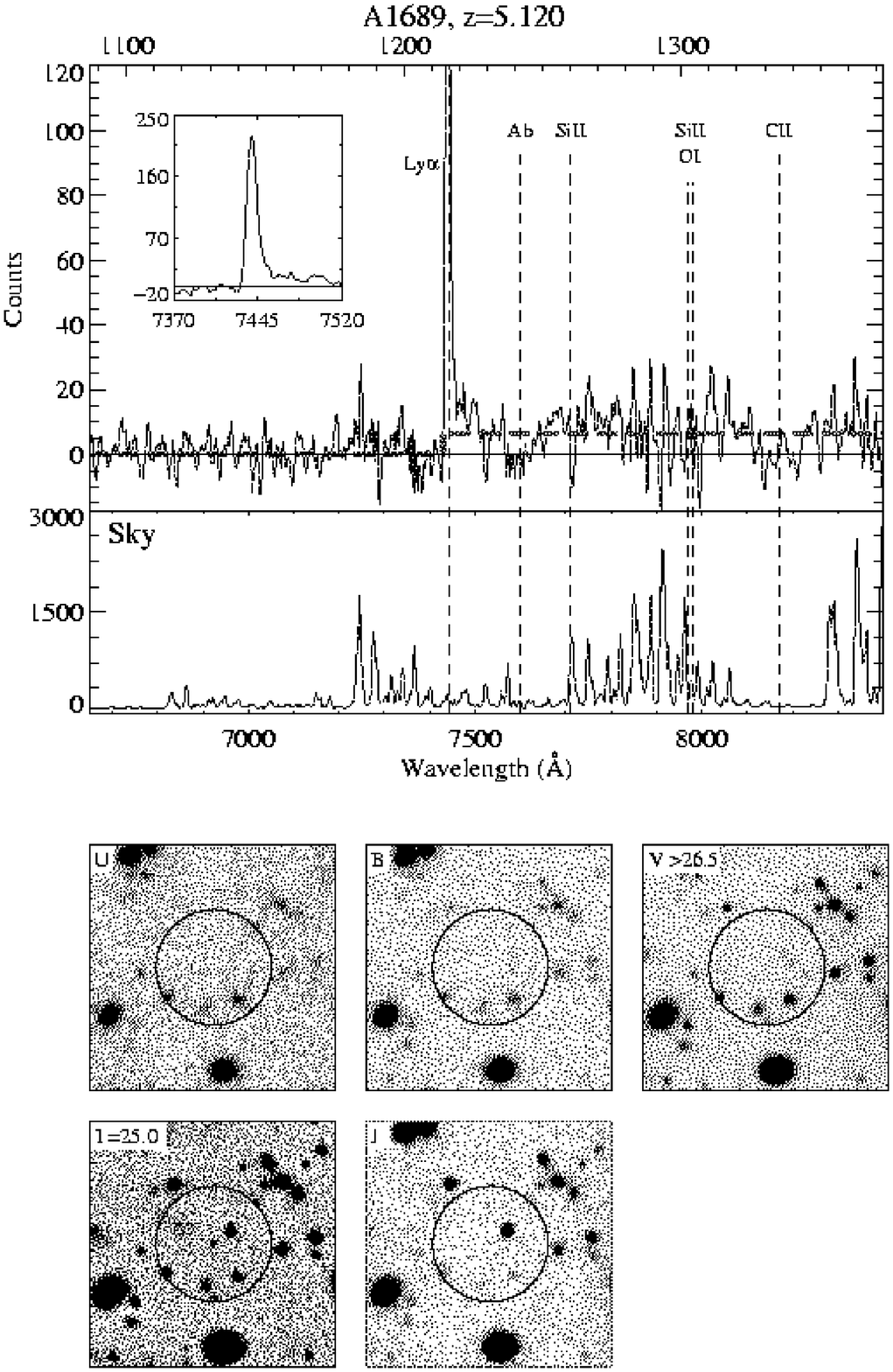}  
\end{figure}  
  
\clearpage  
  
\begin{figure}[tp]  
\plotone{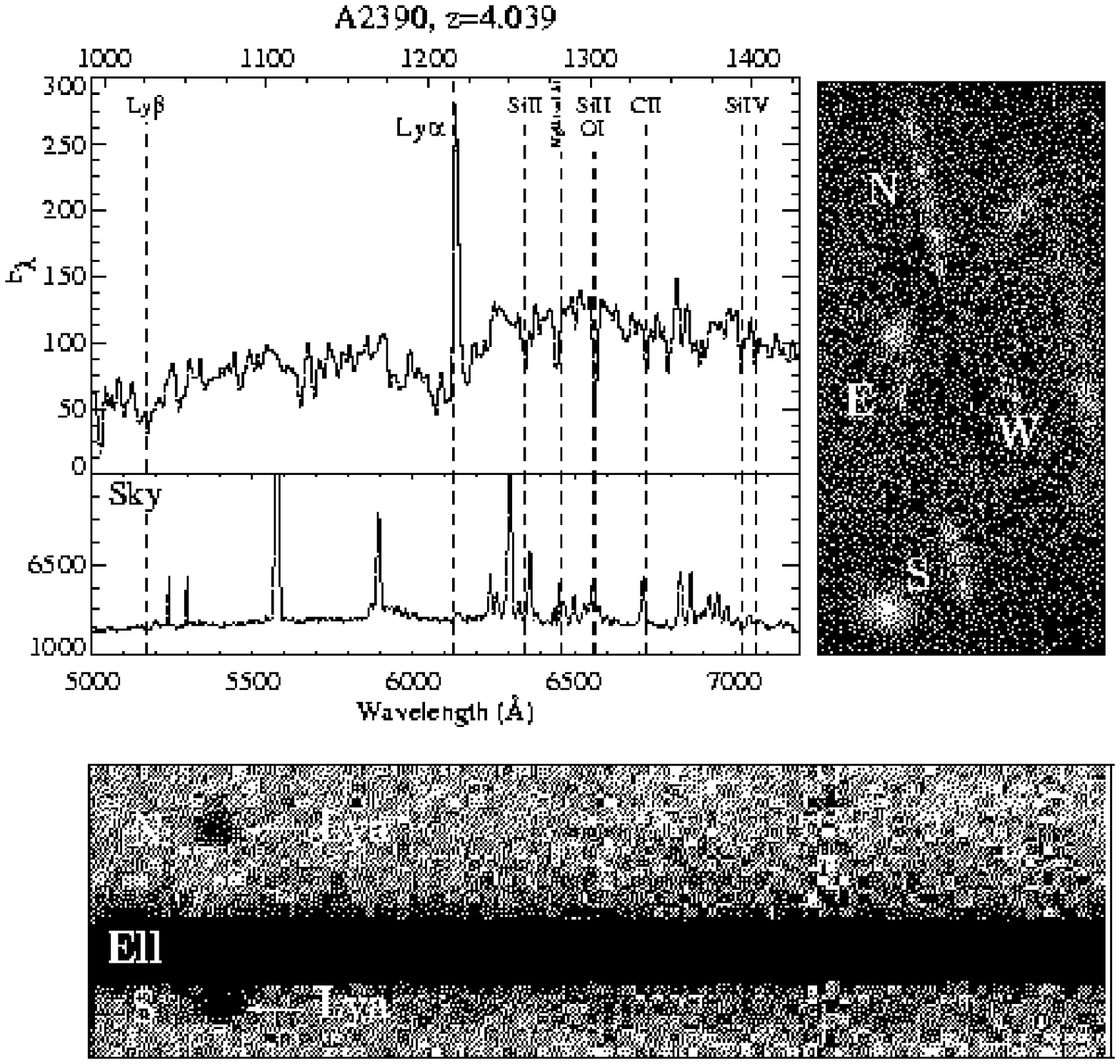}  
\end{figure}  
  
\clearpage  
  
\begin{figure}[tp]  
\plotone{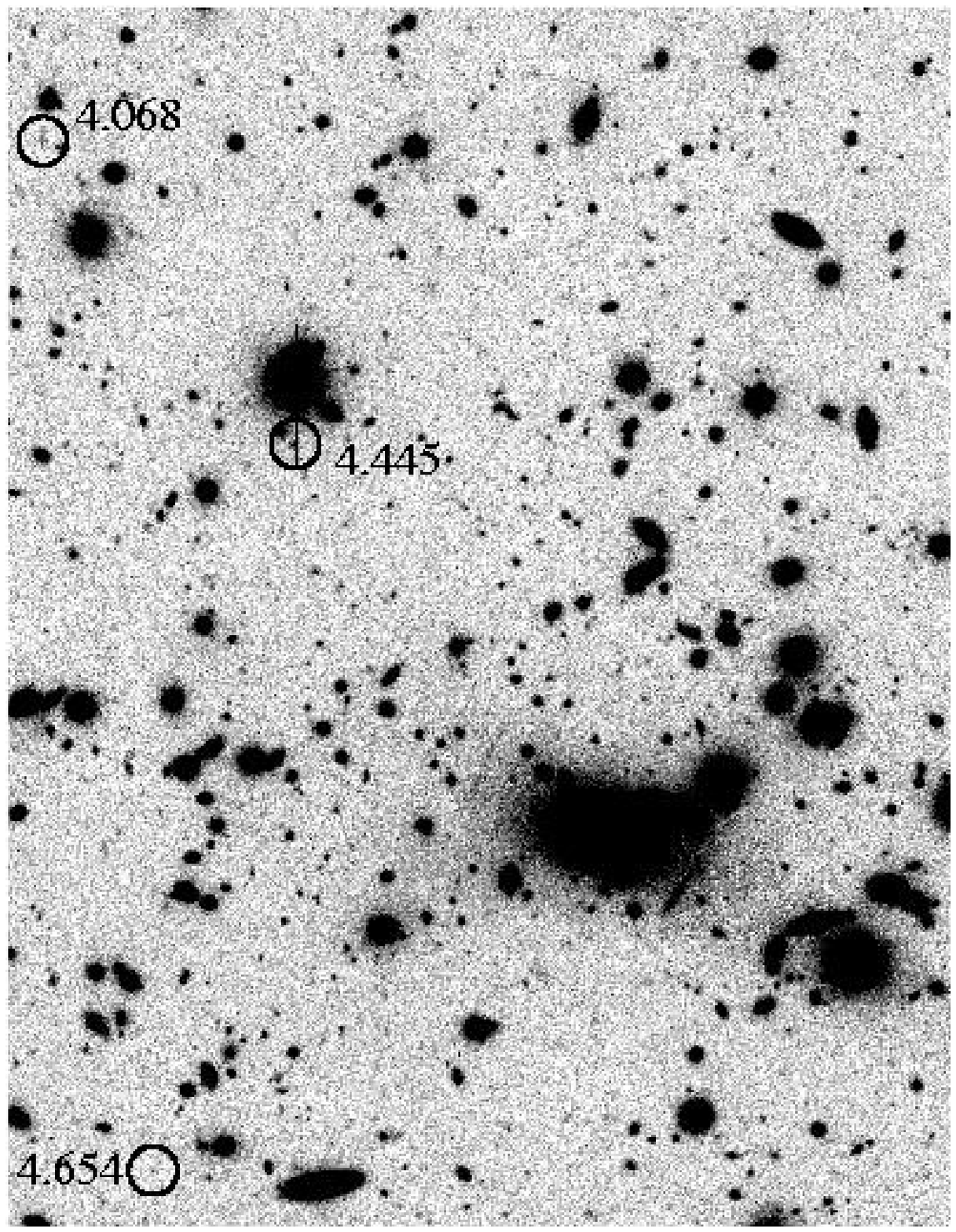}  
\end{figure}  
  
\clearpage  
  
\begin{figure}[pt]  
\plotone{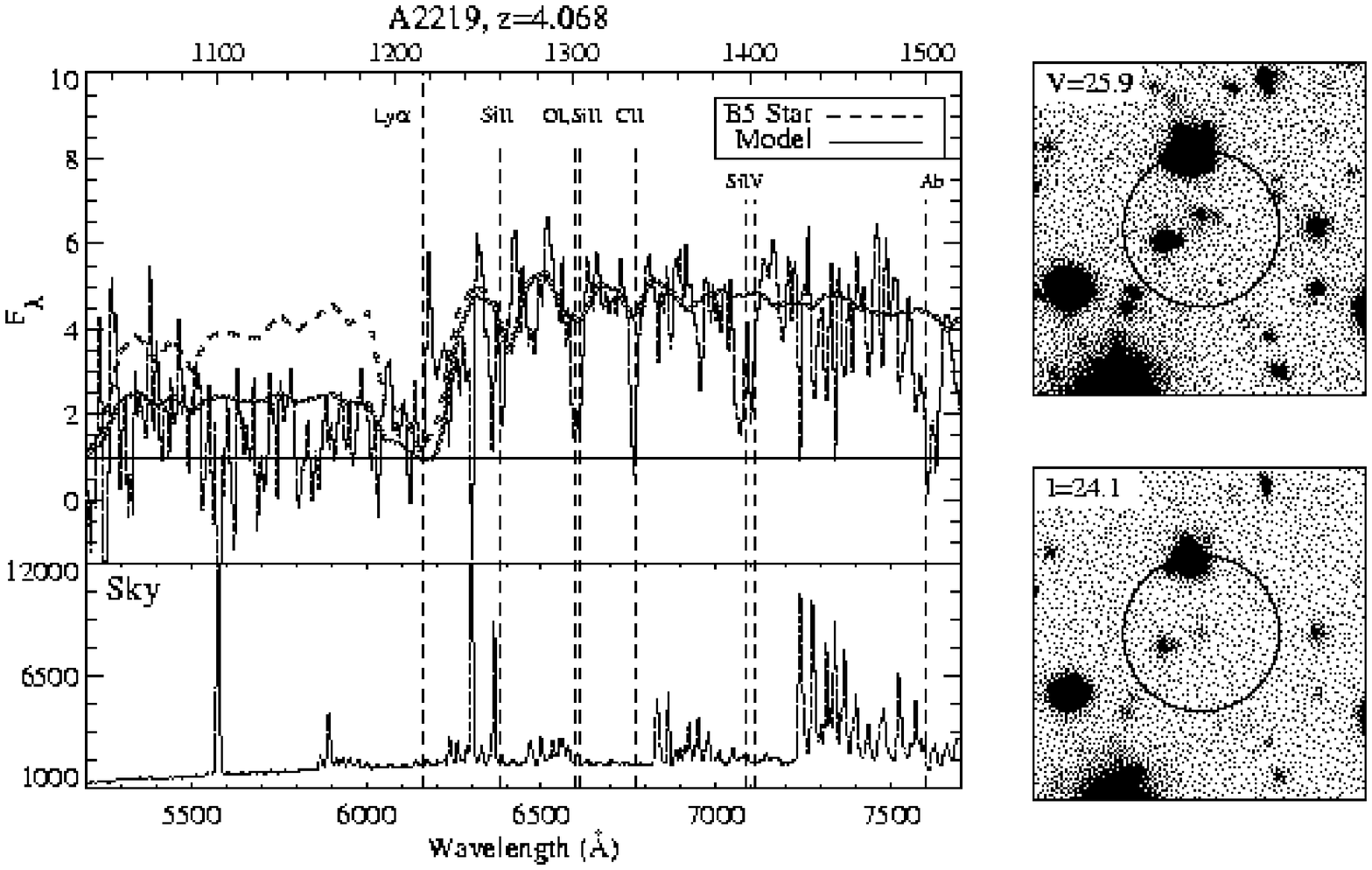}  
\end{figure}  
  
\clearpage  
  
\begin{figure}[pt]  
\plotone{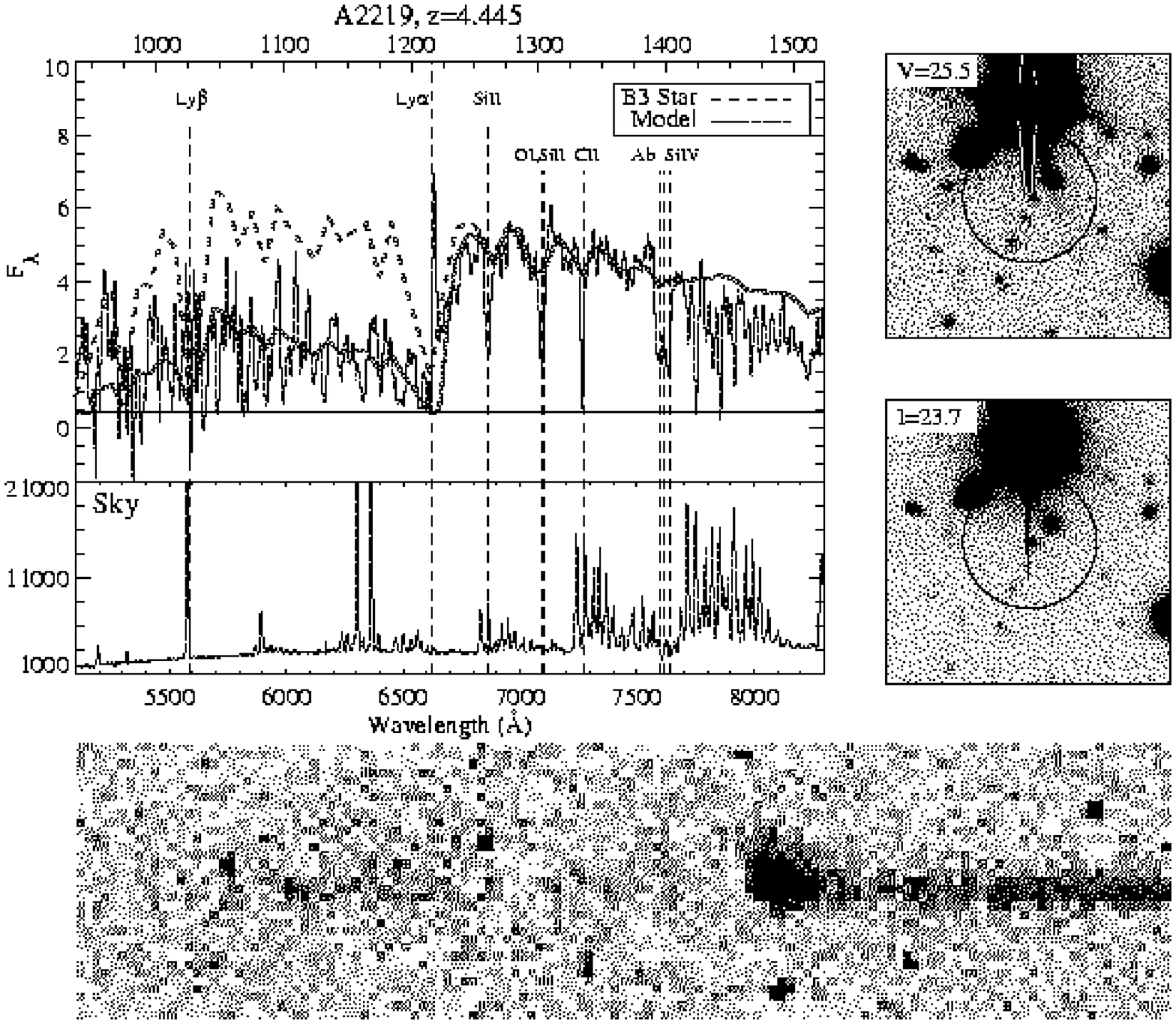}  
\end{figure}  
  
\clearpage  
  
\begin{figure}[pt]  
\plotone{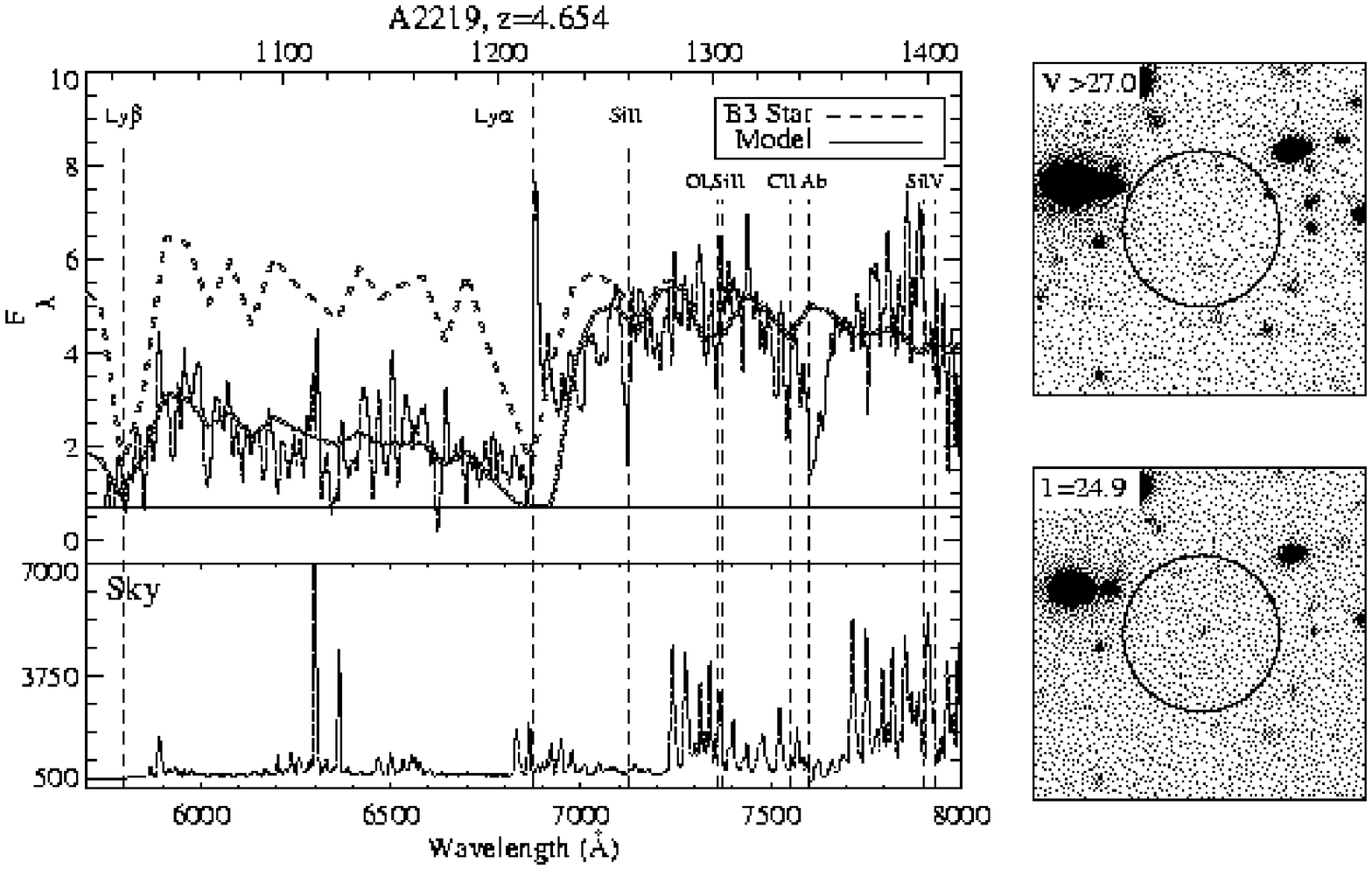}  
\end{figure}  
  
\clearpage  
  
\begin{figure}[pt]  
\plotone{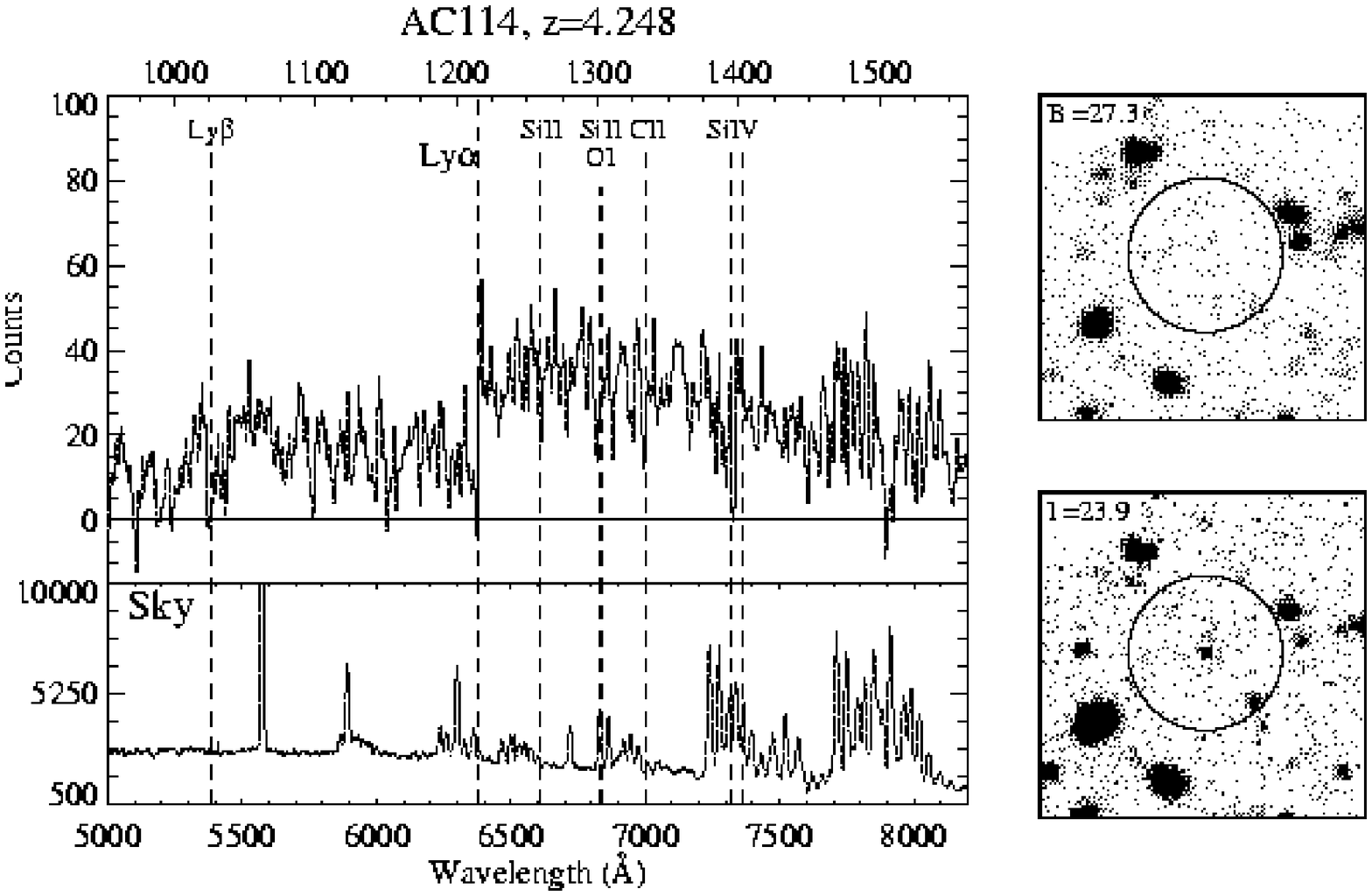}  
\end{figure}  
  
\clearpage  
  
\begin{figure}  
\plotone{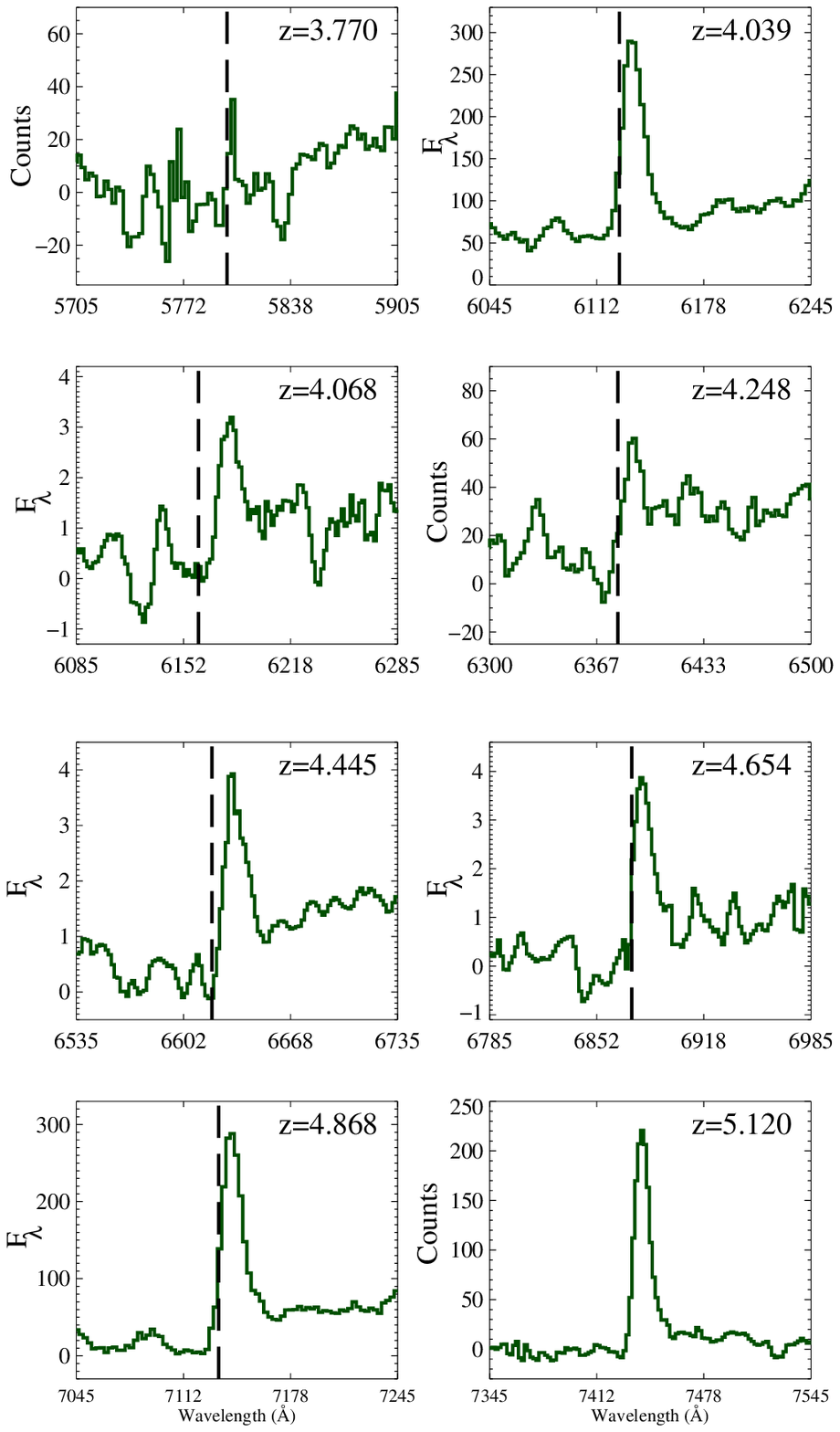}  
\end{figure}  
  
\clearpage  
  
\begin{figure}  
\plotone{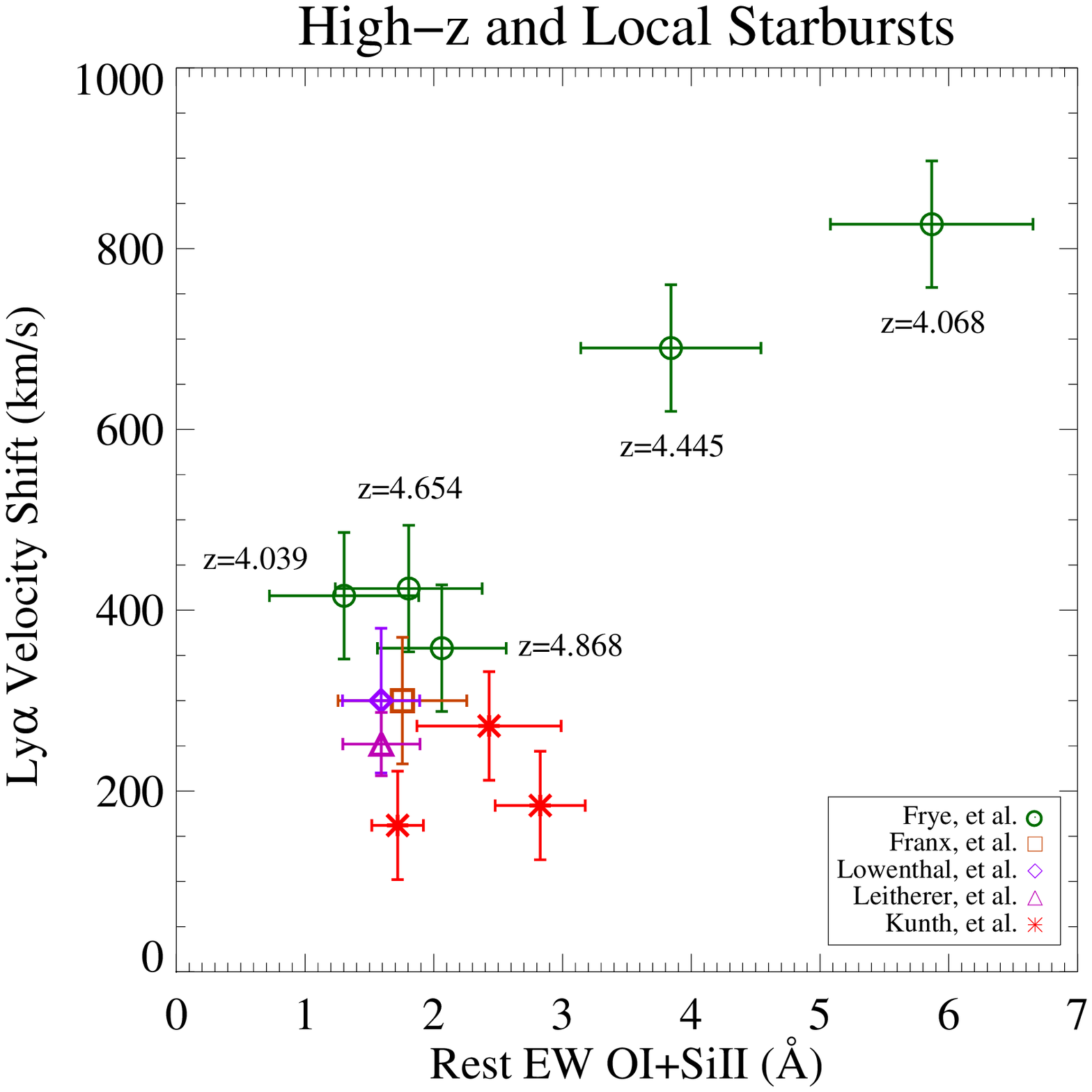}  
\end{figure}  
  
\clearpage  
  
\begin{figure}  
\plotone{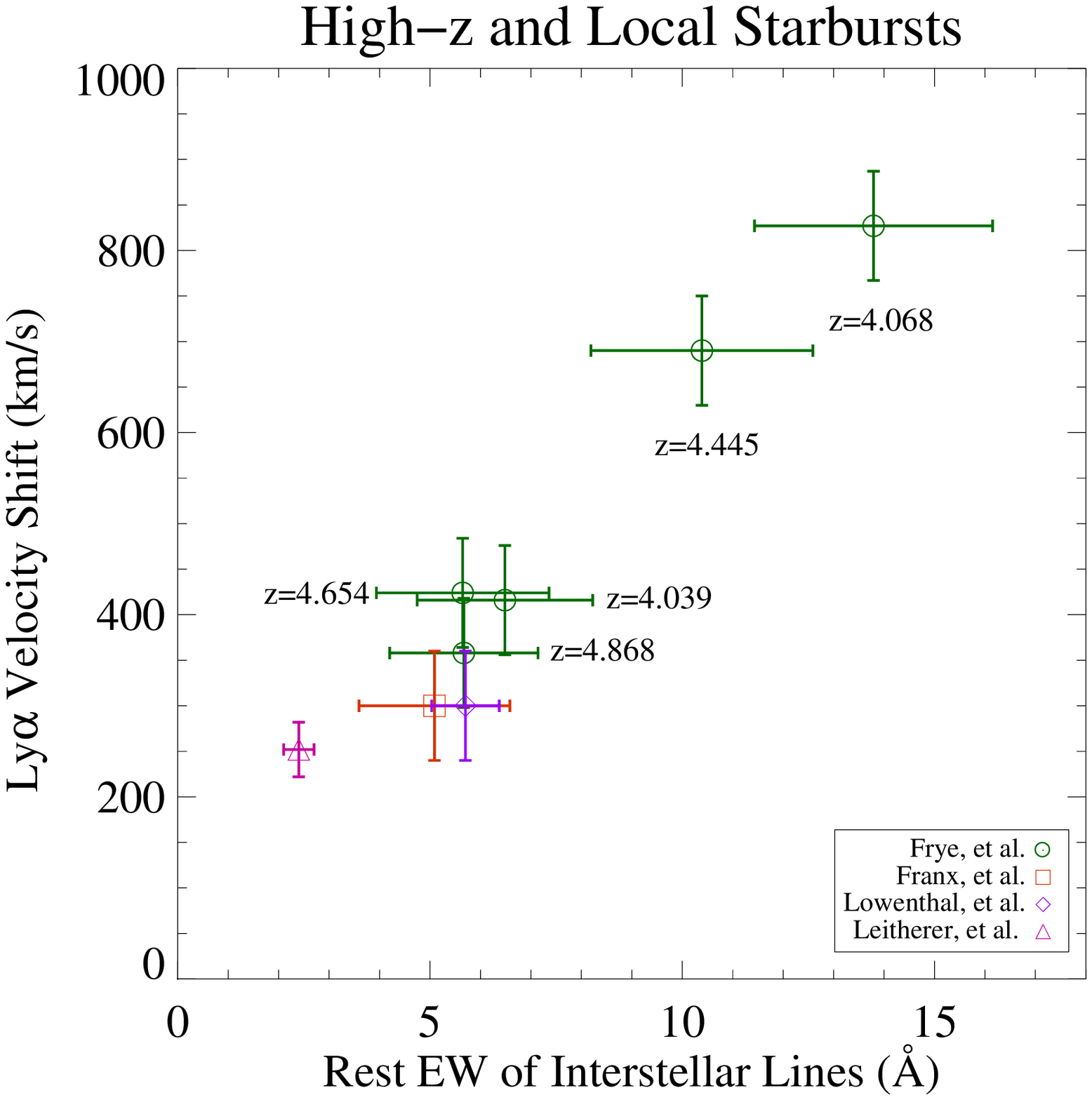}  
\end{figure}  
  
\clearpage  
  
\begin{figure}  
\plotone{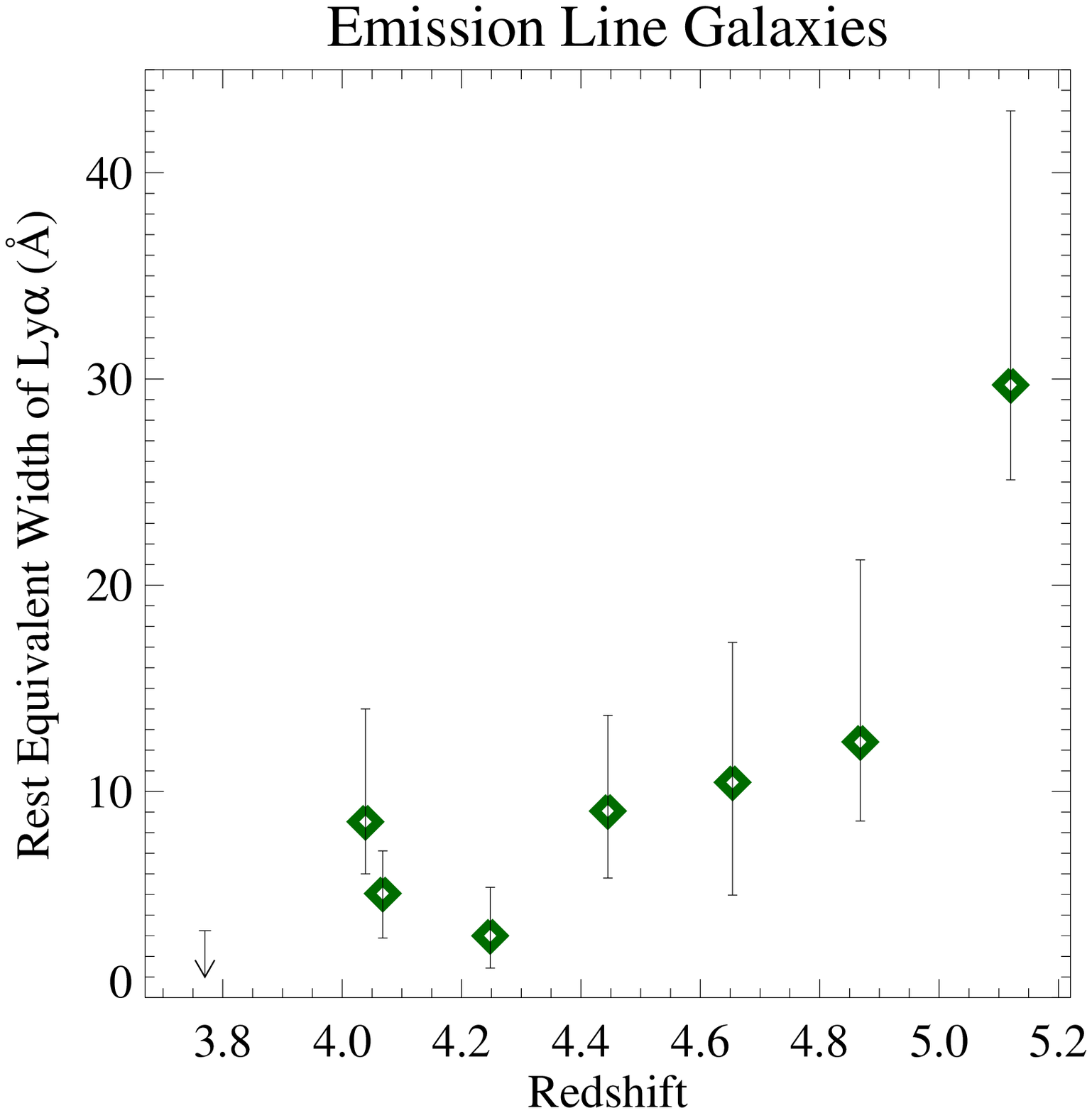}  
\end{figure}

\end{document}